\documentclass[]{aa}  
\usepackage{graphicx}
\usepackage{rotating}
\usepackage{txfonts}
\usepackage[colorlinks=true,linkcolor=blue,citecolor=blue,filecolor=blue,urlcolor=blue]{hyperref}
\usepackage{subfigure}
\usepackage{longtable}
\usepackage{pdflscape} 

\begin{document} 
   \title{Ultra-diffuse Galaxies in the Kilo-Degree Survey with Deep Learning}
   \author{Hao Su
          \inst{1,2}
          \and
          Rui Li\inst{3}
          \and
          Nicola R. Napolitano\inst{1,4,5}\fnmsep\thanks{Corresponding authors: liruiww@gmail.com, nicolarosario.napolitano@unina.it}
          \and
          Zhenping Yi\inst{2}
          \and
          Crescenzo Tortora\inst{6}
          \and
          Yiping Su\inst{2}
          \and
          Konrad Kuijken\inst{11} 
          \and
          Liqing Chen\inst{3} 
          \and
          Ran Li\inst{12} 
          \and
          Rossella Ragusa\inst{13} 
           \and
          Sihan Li\inst{3} 
          \and
          Yue Dong\inst{7}
          \and
          Mario Radovich\inst{8} 
          \and
          Angus H. Wright\inst{9}
          \and
          Giovanni Covone \inst{1}
          \and
          Fucheng Zhong\inst{1,10} 
          }           
   \institute{Department of Physics “E. Pancini”, University of Naples Federico II, Via Cintia, 21, 80126 Naples, Italy
        \and
        School of Mechanical, Electrical and Information Engineering, Shandong University, \\180 Wenhua Xilu, Weihai, 264209, Shandong, China
        \and
        Institute for Astrophysics, School of Physics, Zhengzhou University, Zhengzhou, 450001, People’s Republic of China
        \and
        INAF – Osservatorio Astronomico di Capodimonte, Salita Moiariello 16, I-80131 Napoli, Italy
        \and
        INFN, Sez. di Napoli, via Cintia, 80126, Napoli, Italy
        \and
        INAF — Osservatorio Astronomico di Capodimonte, Salita Moiariello 16, I-80131 Napoli, Italy
        \and
        School of Mathematics and Physics, Xi'an Jiaotong-Liverpool University, 111 Renai Road, Suzhou, 215123, PR China
        \and
        INAF – Osservatorio Astronomico di Padova, vicolo dell’Osservatorio 5, I-35122 Padova, Italy
        \and
        Ruhr University Bochum, Faculty of Physics and Astronomy, Astronomical Institute (AIRUB), German Centre for Cosmological Lensing, 44780 Bochum, Germany
        \and
        School of Physics and Astronomy, Sun Yat-sen University, Zhuhai Campus, 2 Daxue Road, Xiangzhou District, Zhuhai 519082, China
        \and
        Leiden Observatory, Leiden University, Niels Bohrweg 2, 2333 CA Leiden, The Netherlands
        \and
        School of Astronomy and Space Science, University of Chinese Academy of Sciences, Beĳing 100049, China
        \and
        INAF – Osservatorio Astronomico di Capodimonte, Salita Moiariello 16, I-80131 Napoli, Italy
             }

   \date{}

\abstract 
{Ultra-diffuse Galaxies (UDGs) are a subset of Low Surface Brightness Galaxies (LSBGs), showing mean effective surface brightness fainter than $24\ \rm mag\ \rm arcsec^{-2}$ and a diffuse morphology, with effective radii larger than 1.5 kpc. 
Due to their elusiveness, traditional methods are challenging to be used over large sky areas. Here we present a catalog of ultra-diffuse galaxy (UDG) candidates identified in the full 1350 deg$^2$ area of the Kilo-Degree Survey (KiDS) using deep learning. In particular, we use a previously developed network for the detection of low surface brightness systems in the Sloan Digital Sky Survey \citep[LSBGnet,][]{su2024lsbgnet} and optimised for UDG detection. We train this new UDG detection network for KiDS (UDGnet-K), with an iterative approach, starting from a small-scale training sample. After training and validation, the UGDnet-K  has been able to identify $\sim3300$ UDG candidates, among which, after visual inspection, we have selected 545 high-quality ones. The catalog contains independent re-discovery of previously confirmed UDGs in local groups and clusters (e.g NGC 5846 and Fornax), and new discovered candidates in about 15 local systems, for a total of 67 {\it bona fide} associations. Besides the value of the catalog {\it per se} for future studies of UDG properties, this work shows the effectiveness of an iterative approach to training deep learning tools in presence of poor training samples, due to the paucity of confirmed UDG examples, which we expect to replicate for upcoming all-sky surveys like Rubin Observatory, Euclid and the China Space Station Telescope. }

   \maketitle

\section{Introduction}
With extraordinary advances in deep, high-resolution imaging surveys over the past decades, an increasing number of ground-based and space programs have been started (e.g. Euclid, \citealt{laureijs2011euclid, amendola2018cosmology}, the Dark Energy Survey -- DES, \citealt{sevilla2021dark}) or are approaching operations (e.g. Rubin Observatory/LSST, \citealt{abell2009lsst}, the China Space Station telescope -- CSST, \citealt{zhan2011consideration,gong2019cosmology}). These facilities will provide unprecedented survey depths and image quality, enabling the detection of more numerous and fainter astronomical objects, as well as pushing forward the investigation of the low-surface-brightness side of galaxies.Low Surface Brightness Galaxies (LSBGs) have been studied in great detail for decades \citep{impey1988virgo, de1997dark, hayward2005cosmological, du2015low, greco2018illuminating, tanoglidis2021shadows}. Detailed investigations of their properties have drawn attention to a seemingly distinct subclass of ultra-diffuse galaxies (UDGs), originally identified in the Coma cluster \cite[e.g.,][]{van2015forty}.

UDGs are extended LSBGs, with $g$-band central brightnesses $ \mu_{0}(g)\geq 24$ mag arcsec$^{-2}$ and Milky Way-like effective radii $ R_e\geq 1.5$ kpc (\citealt{{van2015forty}}), but stellar masses $\sim10^{2}$ to $10^{3}$ times smaller than the Milky Way, making them comparable to dwarf galaxies. 
Since their postulation and first characterization, they have been regularly searched in deep imaging data \citep{van2017abundance}, although searches have been mostly concentrated in galaxy clusters e.g. in the Hydra cluster \citep{la2022galaxy} . For instance, the Coma cluster alone was found to have $\sim10^{3}$ UDGs 
\citep{koda2015approximately, yagi2016catalog, zaritsky2018systematically, alabi2020expanded}. 
Other studies have targeted Virgo \citep{mihos2016burrell, boselli2016spectacular, lim2020next}, Fornax \citep{venhola2018fornax}, Hydra \citep{iodice2020first}, Perseus \citep{wittmann2017population, gannon2022ultra}, Abell 2744 \citep{lee2017detection}, and Abell 168 clusters \citep{roman2017spatial}, suggesting that UDGs constitute a common population in high-density environments. However, UDGs have been found to populate also galaxy groups \citep{merritt2016dragonfly, roman2017ultra, bennet2017discovery} and the field \citep{bellazzini2017redshift, prole2019observational, leisman2017almost,borlaff2022euclid, marleau2025euclid}.

Despite their ubiquity, and the systematic studies, including spectroscopy \citep[see e.g.,][]{Chilingarian2019_udg_spec,2024Gannon_spec_rev}, their origin and evolution history remains elusive. Understanding why they are such strong outliers in the typical galaxy scaling relations (especially $\mu_e-R_e$, \citealt{roman2017spatial}) is therefore an important question to address \citep{amorisco2016ultradiffuse, bautista2023ultradiffuse}.

One possibility is that UDG formation is driven by external processes, with the star formation in these 'failed' galaxies having been rapidly quenched at high redshift due to environmental processes in galaxy clusters (\citealt{koda2015approximately}; \citealt{yozin2015quenching}). If so, they should be dominated by dark matter, as it seems to be suggested by the large number of Globular Clusters (GCs) in these systems (see e.g. \citealt{saifollahi2022implications}). {In other scenarios, they are disks transformed by the interaction with dense environments (e.g., \citealt{tremmel2020formation})}, leaving the signature in their morphology (low axis ratio, low Sérsic index) or stellar population (old age, short-lived star formation). Supporting this scenario, large UDGs have been associated with tidal material and interaction with companion systems \citep{toloba2015tidally, bennet2018evidence}. On the other hand, internal processes, such as strong stellar feedback and gas outflows, may have dominated the formation of UDGs in isolated environments, leading to their diffuse stellar distributions  \citep{tremmel2020formation}. Finally, UDGs may be “genuine” dwarf galaxies with standard halo mass and luminosity, but anomalously large sizes, produced by the most rapidly spinning systems (\citealt{amorisco2016ultradiffuse}), or by feedback and outflow expanding both the dark matter and stellar component of dwarf galaxies (\citealt{di2024role}), as possibly suggested by the presence of color gradients (\citealt{liu2017origins}).

It is important to better characterize UDGs for their mass content. Although there are limited works directly measuring their dynamical mass,  
the dark matter fraction of UDGs 
is expected to be as large as $>98\% $ \citep{koda2015approximately}. Thus, depending on the effective abundance of UDGs, they can 
contribute with a significant fraction of the measured dark matter 
in the universe. 
For all these reasons, it is essential to collect larger samples of these systems to fully characterize their structure and internal dynamics as a function of the environment in which they live. Large sky surveys are the natural datasets to collect and explore large samples of them. \cite{bennet2017discovery} developed a new detection algorithm, specifically designed for modern wide field imaging surveys, and found 38 unreported diffuse dwarf candidates, of which seven may be UDG candidates. As mentioned previously, the UDG class is a subset of the LSBG, which means they can be identified within existing LSBG collections (\citealt{yagi2016catalog}).   
However, among all the selection methods, the traditional  procedures for detecting UDGs usually involve softwares like Sextractor \citep{bertin1996sextractor} or MTO \citep{teeninga2016statistical} to extract sources from images, identifying their RA and Dec. Subsequently, other softwares, such as GALFITM \citep{peng2002detailed} or IMFIT \citep{erwin2015imfit}, have been employed for fitting their relevant photometric parameters (e.g. their effective radius and mean effective surface brightness), which have been used to select candidates. From these,
the final sample selection is confirmed via visual inspection \citep{van2015forty, yagi2016catalog, alabi2020expanded,bautista2023ultradiffuse}. 
 
Although traditional methods have yielded certain results, they still have limitations. The primary challenge is that these methods are not specifically designed for identifying UDGs; hence, they extract all sources and from the measurement of the structural parameters, they can isolate UDGs. This poses two levels of problems. First, current and future surveys will detect a number of
sources which can be of the order of billions. Thus, extracting and fitting parameters to all these sources would take immense time and resources. Second, due to the low brightness and diffuse morphology of UDGs, uncertainties on structural parameters can be large and using solely them as a criterion to select UDGs, can produce incomplete collections, as candidates can be lost due to the scatter of the size and surface brightness measurements
\citep{he2020sample, yi2022automatic}. This would make the detection of these galaxies highly inefficient and time consuming on 
vast data volumes as the ones collected from Stage III surveys such as the Dark Energy Survey (DES, \citealt{sevilla2021dark}), the Kilo-Degree Survey (KiDS, \citealt{de2013kilo}), and future Stage IV surveys like the Legacy Survey of Space and Time (LSST, \citealt{ivezic2019lsst}), Euclid Mission and the China Space Station telescope (CSST, \citealt{cao2018testing}). Hence, there is a pressing need for developing end-to-end, automated, and reliable methods for large-scale UDG detections in large datasets, e.g trying to minimise the number of candidates over which to perform structural parameters analyses. 

In this paper, our first aim is to introduce a novel deep learning method to detect UDGs, primarily using multi-colour, high quality, ground-based imaging data and provide a first deep-learning based catalog of high-quality UDG candidates in KiDS. 

In a previous work (\citealt{su2024lsbgnet}), we have introduced a framework for detecting LSBGs, called LSBGnet. 
This is based on a You Only Look Once (YOLO) object detection model \citep{redmon2016you}, previously used for object detections in astronomical images (\citealt{grishin2023yolo}), including LSBGs (\citealt{gonzalez2018galaxy}). The network architecture and image processing methods have been adjusted according to the characteristics of LSBGs, including the incorporation of the Coordinate Attention (CA) mechanism \citep{hou2021coordinate}, {\it gamut} transformation, and mosaic data augmentation etc.
In our first application, we have built the LSBGnet-SDSS model and LSBGnet-DES model using data sets from the Sloan Digital Sky Survey (SDSS) and DES, respectively. These models achieve over 97\% recall and precision on the test sets, demonstrating an excellent performance of the framework for LSBGs' detection \citep{su2024lsbgnet}. 
Due to the flexibility of this tool, we want to specialize the LSBGnet to the detection of UDGs. This new UDG detection network, or UDGnet for short, needs to be trained on real UDG images. In this paper, we present the results of a UDGnet model trained with UDGs from the the KiDS fifth "Legacy" data release \citep{wright2024fifth}, which we dub UDGnet-K. 

The first challenge to face is to build-up an {\it ad hoc} training sample to train the tool for the systematic application to the whole KiDS dataset.
Indeed, only a few UDGs are known in KiDS from \citet{van2017abundance}, which has performed a systematic detection for UDGs in a selected sample of galaxy groups, in 197 deg$^{2}$ 
covered by Galaxy and Mass Assembly (GAMA, \citealt{driver2011galaxy})\footnote{Unfortunately, no catalog was attached to the paper.}. Hence, any training sample one can assemble will be necessarily smaller than standard samples usually adopted for training Deep Learning methods (i.e. Convolutional Neural Networks -- CNNs). Thus, we will first focus on finding a strategy to enhance the training process, starting from a small training set, and then we will effectively and accurately perform a first selection of the UDG candidates in the full KiDS DR5, covering an area of $1350 $ deg$^{2}$ area \citep{wright2024fifth}.  

For this specific work, we are not interested in fully quantifying the photometric properties of these candidates via surface brightness modeling, as the main purpose is to provide an automated method to robustly identify UDGs in imaging data. Visual inspection of expert observers will be the main step to score the quality of the deep learning detection. However, we will still produce some realistic structural quantities to validate and check for visual biases.

This paper is organized as follows: in Section \ref{sec:data}, we describe the KiDS data and the processing of the data set. In Section \ref{sec:method}, we illustrate the detection process of UDG candidates. In Section \ref{sec:results}, we present the specific detection procedure and the corresponding results. Section  \ref{sec:Final Sample Selection} focuses on further filtering of the detected candidates to obtain the High-quality sample and an analysis of their spatial distribution. Finally, in Section \ref{sec:Summary}, we summarize our work and draw the conclusions.

\section{Data} \label{sec:data}

\begin{figure}
    \centering
    \includegraphics[width=0.48\textwidth]{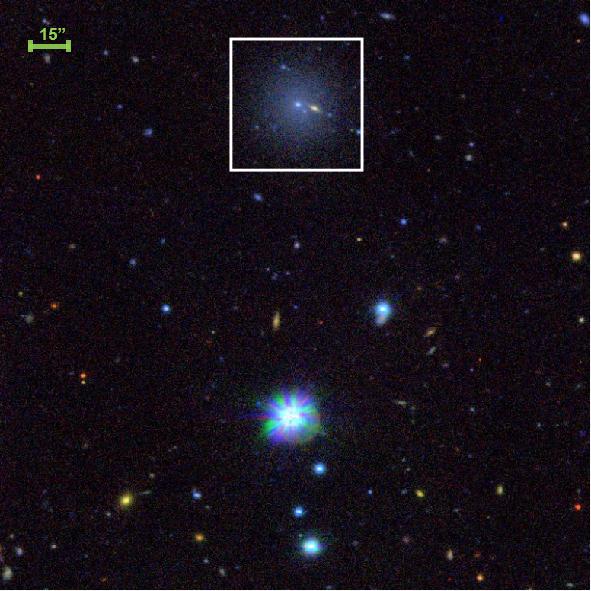}
    \caption{The gri composite KiDS image with the white box highlighting a  UDG candidate detected in the initial sample.}
    \label{fig1}
\end{figure}

KiDS is a Stage III optical wide-field imaging survey \citep{de2017third}, carried out at the VST telescope \citep{capaccioli2011vlt} with the OmegaCAM camera \citep{kuijken2011omegacam}, located at the European Southern Observatory (ESO), Cerro Paranal Observatory, in Chile. The final Data release 5 (DR5) of the full area of $\sim1350$ deg$^{2}$, observed in four optical filters ($ugri$), reach a 5$\sigma$ point-source depth of $r \sim 25.1$ mag and a surface brightness limit of $\mu_e \sim 26.4$ mag arcsec$^{-2}$ \citep{{wright2024fifth}}.
In the DR5, as done in previous releases, the optical imaging from VST has been combined with Near Infrared (NIR) data of the VISTA Kilo degree Infrared
Galaxy (VIKING, \citealt{edge2013vista, venemans2015first}), which have observed the same KiDS area in ZYJHK using the Visible and Infrared Survey Telescope for Astronomy (VISTA), also located in Cerro Paranal. The total KiDS-DR5 then provides a unique 9-band multicolor data set. 
Being KiDS primarily designed to map the large-scale distribution of matter in the Universe via weak lensing, it is characterized by a very high image quality ($\sim0.7''$ in the $r$-band) that, combined with the dark sky of Paranal, makes the KiDS data ideal for low-surface brightness studies \citep{2018roy_kids_sersic,2018kelvin_surface_bright_kids}.  

\begin{figure*}[ht!]
    \centering
    \includegraphics[width=0.9\textwidth]{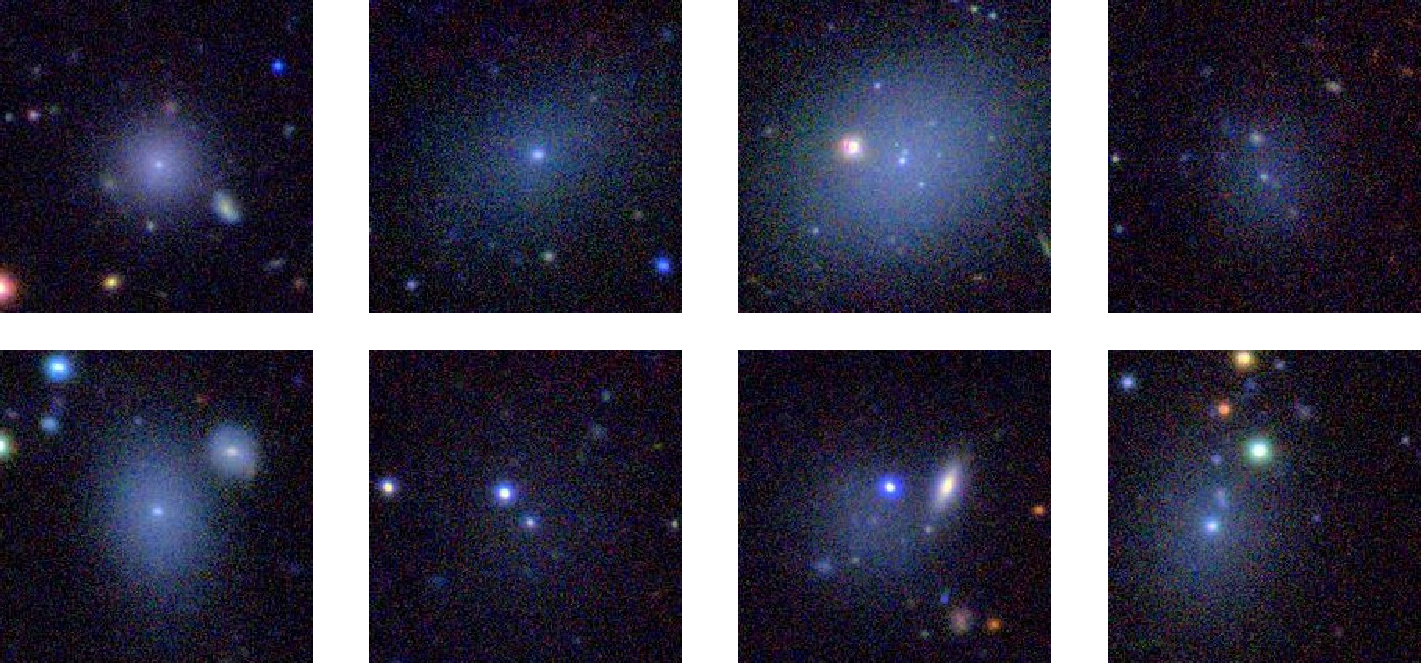}
    \caption{Images of some UDG candidates in the training set. The size of each image is 40$'' \times 40''$.}
    \label{fig2}
\end{figure*}

In this paper, we intend to search for UDG candidates using $gri$-band composite images, consisting of large cut-outs, or ``chunks'', of the sky. To ensure compatibility with the input dimensions of the UDGnet framework, we use chunks of 1001$\times$1001 pixels, corresponding to a field of view of approximately 3.33' at a pixel scale of 0.2 arcsec/pixel. An example of a sky chunk with a UDG candidate on top is shown in Fig. \ref{fig1}. 
Given the large effective radii of UDGs, to minimize mis-detections due to being too close to the edge of a sky chunk, we set a 100 pixel ($ \sim 20''$) overlapped among the different chunks. After the ``cutting out'' process, we have obtained a total of 592,620 images completely covering the full KiDS area. 

As we intend to use a deep learning-based object detection algorithm (see \S\ref{sec:method}), this will require us to provide not only the class of the objects, but also their location and size, indicated by a bounding box. In this work, we used LabelImg software \citep{tzutalin2015labelimg} to manually label the UDG objects.

\section{Methods} \label{sec:method}
In this section, we illustrate the deep learning model adopted to obtain the detection of the best UDG candidates.  
The iterative process is implemented by augmenting the training sample, based on the step-by-step UDG detection, until the training sample has reached a sufficient size to proceed to the final selection of the best UDG candidates.
We also introduce the definition of the structural parameters that will be used to photometrically characterize the UDG candidates.

\subsection{The UDG detection model (UDGnet)}
\label{sec:model}
As mentioned, the UDGnet is based on the previously developed LSBGnet framework. The LSBGnet framework consists of four main steps: 1) Image data augmentation; 2) Building a LSBGs detection network; 3) Defining a loss function; 4) Optimizing the network parameters and improving the model performance through iterative learning. Here, we did not update the detection network; we modified the image data augmentation based on the characteristics of the training images and UDGs. Additionally, we employed an iterative detection training strategy to build the model in the subsequent process.

The philosophy behind the UDGnet, equally to the original LSBGnet, is data-driven, thus the characteristics of the training data largely determine the final detection performance of the model. Specifically, the images in the training set first underwent built-in data augmentation. 
This consists on scaling, flipping, color {\it gamut} transformation and mosaic data augmentation. 
The augmented images are overlaid onto a 1024$\times$1024 pixels gray-scale image before being input into the model. In previous works, most galaxy samples were located centrally within images, which increased the risk of overfitting and decreased model robustness. To address this, we incorporate the enhancement of the Mosaic data to improve the generalization of the model \citep{su2024lsbgnet} and avoid to locate 
the UDG images in a central position. 
To streamline the training process and reduce computational overhead, we limit the proportion of mosaic data augmentation to 10\% in the UDGnet framework. We also modified the range of scaling and {\it gamut} transformation data augmentation to ensure that the augmented images better align with the characteristics of UDGs.

\subsection{The UDGnet for KiDS (UDGnet-K)}
\label{sec:udgnet-k}

Finally, the new UDGnet also differs from the previous LSBGnet for the training set, which needs to be built on a specific high-quality candidates, preserving the same noise, depth and image quality of the images we need to use for the detection. 

Here, we present the strategy adopted for the  UDGnet-K, trained on KiDS images. As anticipated, the UDGnet training is based on an iterative approach. This starts with the initial step of building-up the training set, a small initial sample of UDG candidates using visually inspecting a randomly selected area of $ \sim30 $ deg$^{2}$ from the KiDS sky region, comprising 15,000 images. 
We identified a sample of 35 UDG candidates, some of which 
are shown in Fig. \ref{fig2}. 
Looking at the sample in Fig. \ref{fig2}, we see that the main qualitative criterion in the UDG visual selection is the uniform light distribution with little or no substructure. In the first place, we have excluded LSB systems with plumes or knots, because such substructures are not usually seen in UDG candidates, except possibly in the field (see e.g. \citealt{prole2019observational}). We will discuss later about the impact of this choice. For the moment, we remark that if, on the one hand, regular and smooth profiles represent the majority of the confirmed UDGs (see e.g. \citealt{roman2017ultra, koda2015approximately}), on the other hand, this does not imply that the UDGnet will select only such kind of systems, as the presence of background systems can still mimic the presence of substructures in the training sample.

\begin{figure}
    \centering
  \begin{subfigure}{}
    \includegraphics[width=0.49\textwidth]{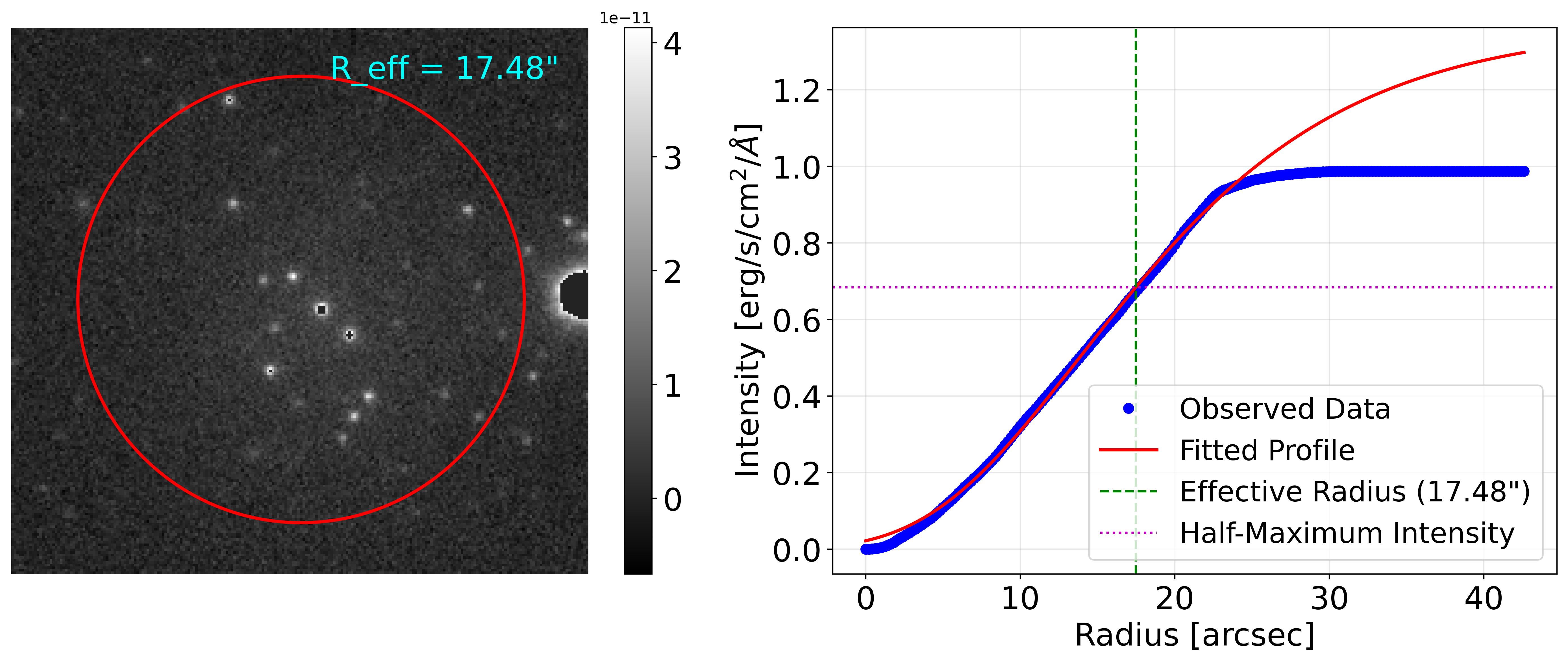}
    \label{Figure3(a)} 
    \includegraphics[width=0.49\textwidth]{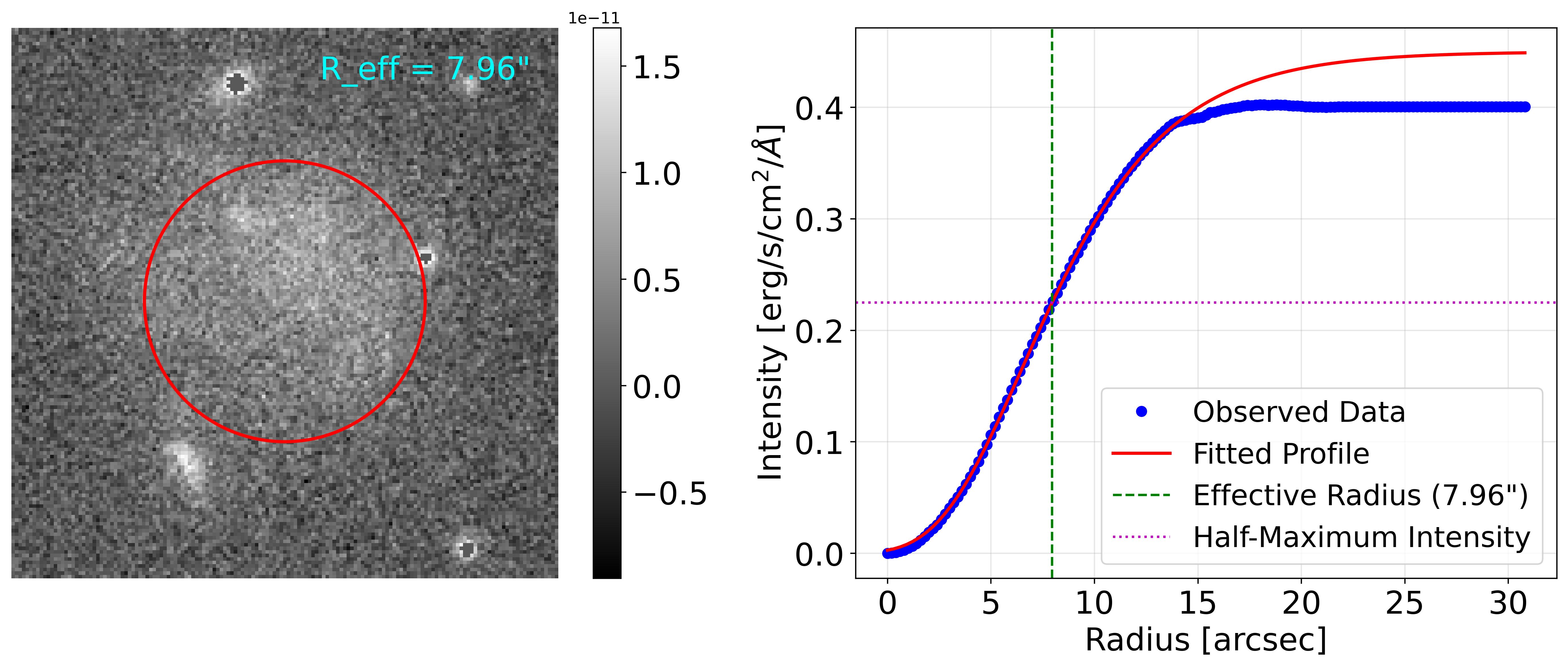}
    \label{Figure3(b)} 
    \includegraphics[width=0.49\textwidth]{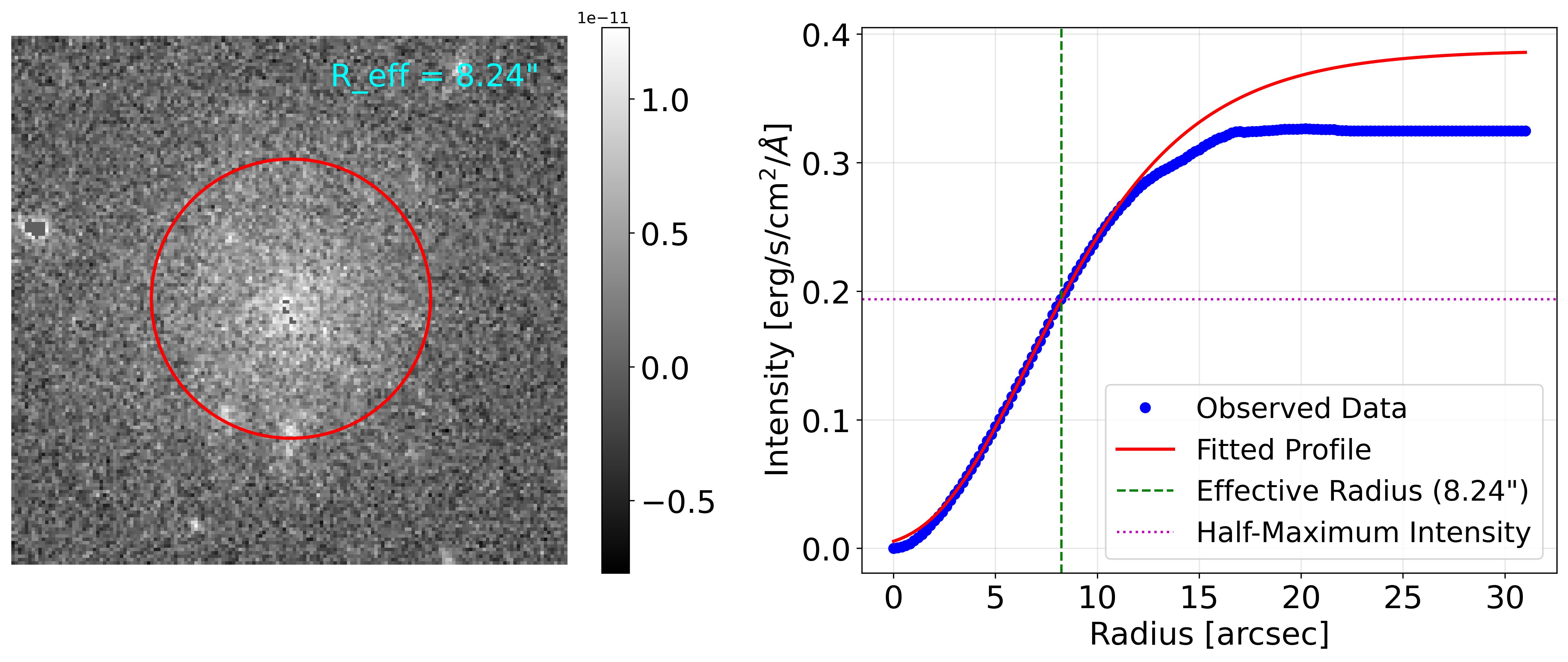}
    \label{Figure3(c)} 
  \end{subfigure}
    \caption{The ``growth curve'' for a UDG candidate. In each of the left panel, the red circle represents the effective radius determined as the radius that encloses half of the total light, obtained by the growth curve in the right panel. The masks applied to exclude the brightest background/foreground objects are also shown in the left columns. The first row shows a known UDG candidate in the NGC 5846 galaxy group.}
    \label{fig11}
\end{figure}

\subsection{Structural Parameters} \label{sec:Analysis of the candidates}
\label{sec:struc_par}
To characterize the candidates identified by the UDGnet-K model, we need to compute their surface brightness and effective radius. As mentioned earlier, in this paper we will not perform a full surface brightness fitting , e.g. using a S\'ersic profile (see \citealt{van2017abundance}). 
However, we will introduce a simpler estimate of the effective radius, $R_e$, and the brightness of the mean effective surface, $\langle \mu_e\rangle_r$, based on the luminosity growth curve in the $r$-band. We use the $r$-band because this is the highest quality band in KiDS (FWHM$\sim0.7''$, see \citealt{kuijken2019fourth}), hence we expect this to provide more unbiased estimates of the structural parameters in which we are interested. As anticipated,  we  stress here that the seeing of the $r$-band images is smaller than the smallest effective radii we will consider for the UDG candidates (i.e. 2$''$, see \S\ref{sec:reff}), hence we expect the seeing lightly affecting the $R_e$ inferences, especially for large-angular systems. Finally, since we use $r$-band instead of the canonical $g$-band used to characterize the UDGs, 
we will adopt a different standard for the UDG definition based on the mean surface brightness in $r$-band of $\langle \mu_e\rangle_r\geq 24$ mag/arcsec$^2$, widely used in literature \citep{van2017abundance, pina2019off}.

\begin{figure}[]
    \hspace{-0.5cm}
    \includegraphics[width=0.51\textwidth]{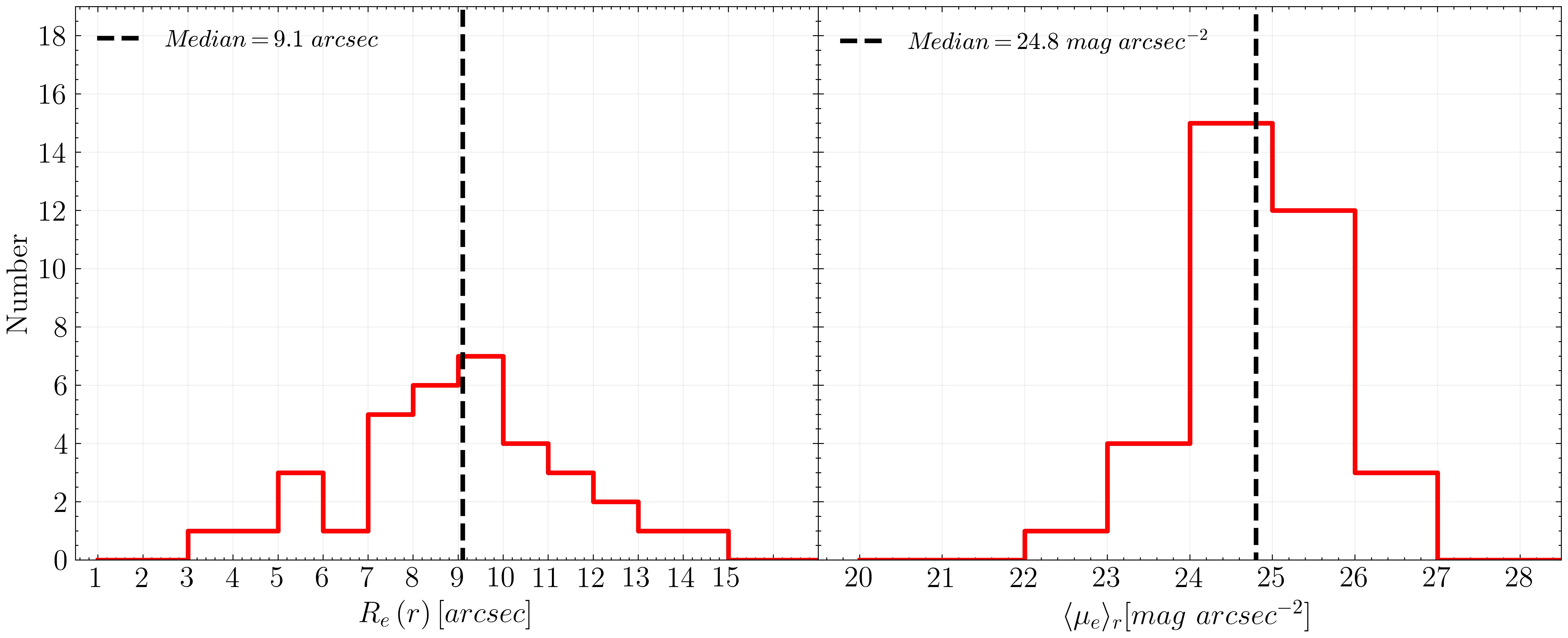}   
    \caption{The distribution of effective radius $R_e$ and mean effective surface brightness $\langle \mu_e\rangle_r$ of the initial sample. The black dashed lines represent the median effective radius and mean effective surface brightness of the UDG candidates.}
    \label{fig4}
\end{figure}

\begin{figure*}
  \centering
  \begin{subfigure}{}
  \includegraphics[width=0.45\textwidth]{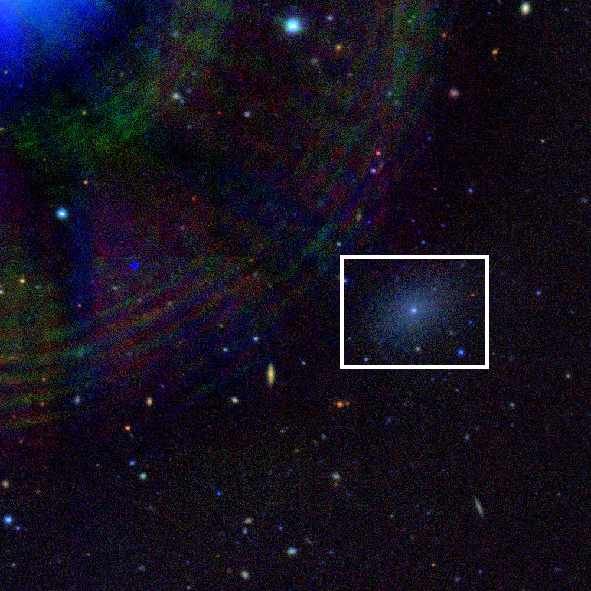}
  \end{subfigure}
  \begin{subfigure}{}
  \includegraphics[width=0.45\textwidth]{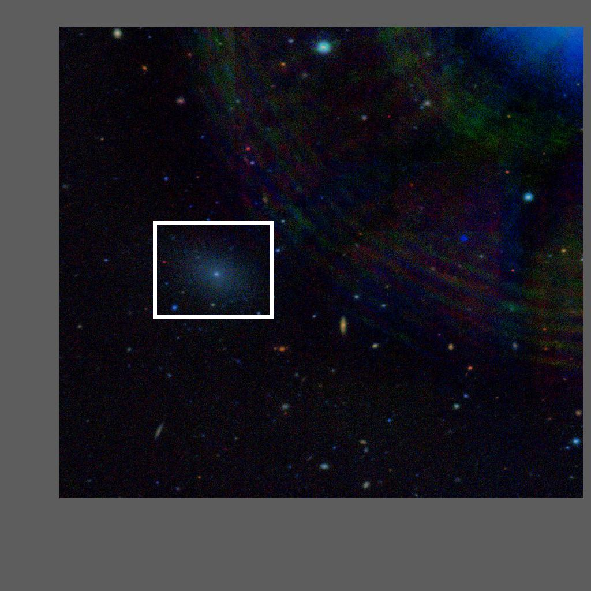}
  \end{subfigure}
  \caption{The comparison of images before and after data augmentation, the panel on the left displays the original photometric image in the initial sample, and the right one shows the image after data augmentation. The white box indicates the UDG object.}\label{figure3}
\end{figure*}

Before we detail the measurement of structural parameters, we stress that while the surface brightness is a distance-independent quantity and can directly be used to characterize the ``diffuseness'' of the candidates, the $R_e$ needs a distance to be converted on the linear scale from the angular scale is measured on the images. As we will detail later (see \S\ref{sec:matching_kids}), even if the UDGnet-K candidates have correspondent sources in the 
KiDS catalogs, their photometry is expected to be highly uncertain due to the intrinsically low surface brightness of these systems, making the publicly released photometric redshifts unreliable (\citealt{de2015first}). Moreover, these objects are expected to lie at low redshift (likely $z < 0.1$), a regime where photometric redshifts are generally less precise, further limiting their utility for accurate distance determinations.

\subsubsection{Effective radius}
\label{sec:reff}
For the measurement of the effective radius we start by adjusting the bounding boxes predicted by the UDGnet-K model to ensure they completely encompassed the UDG candidates, including their surrounding diffuse halos. Next, due to the presence of field stars and galaxies overlapping with the surface brightness extension of the UDGs, we build a mask to exclude all pixels with fluxes exceeding ten times the average flux within the adjusted bounding box region. After masking these bright objects, we calculate the cumulative luminosity within circular areas with increasing values of their radii, $L(<R)$. Importantly, we avoid using the outer regions where the growth curve flattens, as these areas may be affected by background over-subtraction that could lead to systematic underestimation of $R_e$. To ensure a robust measurement, we instead derive $R_e$ from a parametric fit to the inner, well-constrained portion of the growth curve. We model the luminosity growth using a double-exponential function. First, we convert the surface brightness profile $\mu(R)$ into linear intensity units ($ \text{erg/s/cm}^2/\text{\AA} $):
\begin{equation}
\label{eq1}
     Flux=\frac{10^{-0.4\mu(R)}}{10^{-7}}
\end{equation}
where the $ 10^{-7}$ term normalizes the flux scale. The growth curve is then fitted with:
\begin{equation}
    \label{eq2}
    L(R) = L_{\text{max}} \, \exp\left(-\exp\left(-k(R - R_0)\right)\right)
\end{equation}
Here $L_{\rm max}$ represents the asymptotic total luminosity, $k$ controls the steepness of the curve, and $R_{0}$ marks the transition radius where the growth rate changes. This functional form provides a flexible fit to the observed profile while minimizing sensitivity to noise in the outskirts. The effective radius $R_e$ is derived from the best-fit model, corresponding to the radius enclosing half of $L_{max}$, and the best-fit curve is shown in Fig. \ref{fig11}. In Fig. \ref{fig4}, we show the distribution of the derived $R_e$ for the visual sample of 35 UDGs, discussed in \S\ref{sec:model}. As we can see they are mostly distributed in the range of $3-15$ arcsec, consistently with what was found by \citet{van2017abundance}. However, systematic studies of UDG candidstes (e.g. \citealt{zaritsky2023systematically}) have found even broader effective radius distributions, reaching $R_e=20''$. In the following, we will use the range $3-20$ arcsec as a fiducial interval for realistic UDGnet candidate sizes. Typical errors on the estimate of these effective radii have been evaluated by varying the upper data-point to be used to fit the growth curve before reaching the plateau.
By perturbing this upper limit we have found the effective radius estimates to vary up to $\pm$0.05dex in $\log R_e$.

\subsubsection{Effective surface brightness}
\label{sec:central surface brightness}
We have anticipated that, for a galaxy to be defined as a UDG, we adopt 
a lower mean surface brightness $\langle \mu_{e}\rangle_r\geq 24$ mag arcsec$^{-2}$.
The $r$-band $\langle \mu_{e}\rangle$, or $\mu_{e}$ for brevity, is defined as
\begin{equation}
\label{eq3}
    \mu_{\rm e}=-2.5\log_{10}(Flux_e)
\end{equation}
where $Flux_e$ is defined as $Flux_e=L(<R_e)/\pi R_e^2$, where $L(<R_e)$ is the total $r$-band luminosity\footnote{All luminosities derived by the KiDS images have been extinction corrected by averaging the extinction values of close objects to the coordinates of the candidates in the KiDS catalog.} within the effective radius $R_e$
obtained from the growth curve $L(<R)$ as in \S\ref{sec:reff}, and 
$\pi R_e^2$ is the area enclosed by $R_e$.

As for the effective radii, by changing the fitting upper limit of the growth curve, we have also estimated the typical error on $\langle \mu_{e}\rangle_r$ and found it to be of the order of 0.2 mag/arcsec$^2$. To account for these errors and minimize the lost of good candidates that migh fall out of the definition because of errors on $\langle \mu_{e}\rangle_r$, we have decided to finally adopt a lower conservative limit of $\langle \mu_{e}\rangle_r>23.8$ mag/arcsec$^2$.
In the right panel of Fig. \ref{fig4} we show the distribution of the $\langle \mu_{e}\rangle_r$ of the visual sample of 35 UDGs. We see that most of the estimated $\mu_e$ are indeed compatible with the $r$-band 23.8 mag/arcsec$^2$ as a lower limit for being UDGs. Only 4 of them turn out to have $\langle \mu_{e}\rangle_r<23.8$ mag/arcsec$^2$, which we have excluded to finally retain 31 objects with $r$-band $\langle \mu_{e}\rangle_r>23.8$ mag/arcsec$^2$.

\subsubsection{Initial sample}
\label{sec:initial_sample}
Here, we summarize the properties of the initial sample to be used in the next steps of the UDGnet-K training, validation and results. This is the critical starting point as, depending on the criteria adopted, we might produce a biased UDG candidates' catalog. This is made up of 31 candidates visually selected from a KiDS a randomly selected area of $\sim30$ deg$^2$, whose properties are:
\begin{enumerate}
\item growth curve circular effective radius in the range $3''<R_e<20''$, consistently with \citet{zaritsky2023systematically};
\item $r$-band mean effective surface brightness $\langle \mu_{e}\rangle_r>23.8$ mag/arcsec$^2$, considering the typical error on growth curve.
\end{enumerate}

\begin{figure}[]
    \hspace{-0.5cm}
    \includegraphics[width=0.48\textwidth]{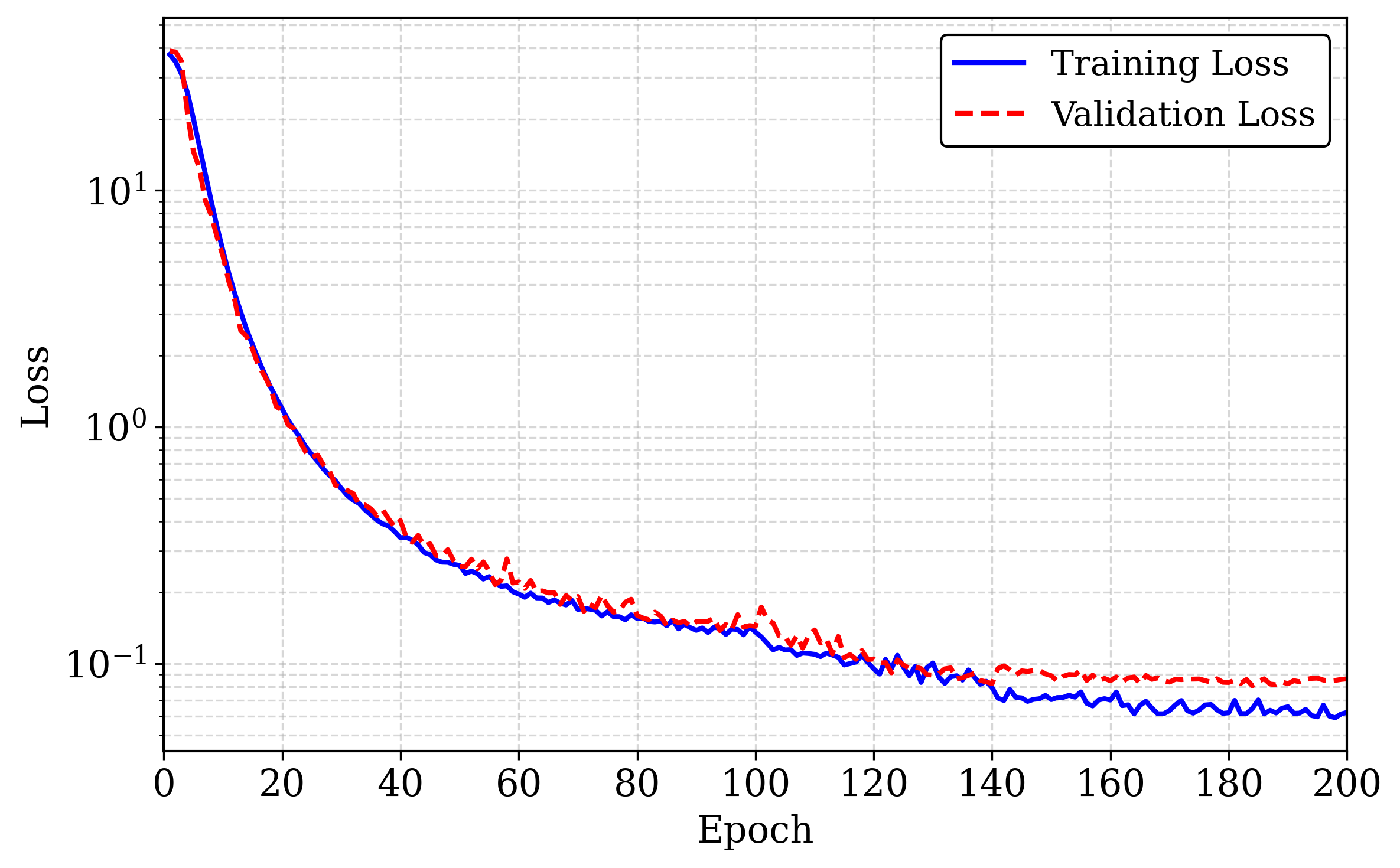}   
    \caption{The loss curves of training and validation.}
    \label{loss}
\end{figure}

\section{Results}\label{sec:results}

\subsection{Sample Expansion} \label{subsec:Sample Expansion}

The use of an adequate training sample is crucial for the UDGnet framework to effectively learn the features of the objects. To enable UDGnet-K to capture key characteristics, such as the morphology and brightness of the UDGs in the KiDS images, we use data augmentation to expand the initial sample obtained in Section \ref{sec:initial_sample}.

In particular we have applied scaling, flipping, and color {\it gamut} transformation. The comparison of the images before and after data augmentation is illustrated in Fig. \ref{figure3}. 
{Here we can see that UDG candidates, after the data augmentation (scaling, flipping, and color gamut transformation), still retain their characteristic LSB and diffuse morphology. Through data augmentation, we expanded the initial sample of 31 {\it initial} UDGs to a total of 100, more than tripling the original sample size.}

\begin{figure}
    \centering
    \includegraphics[width=0.44\textwidth]{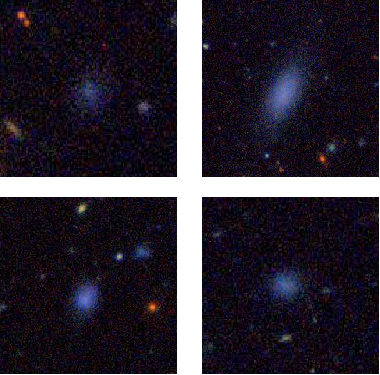}
    \caption{Images of the four candidates detected in the first round test set of the iterative training set building-up.}
    \label{fig6}
\end{figure}

\subsection{Iterative Detection} \label{subsec:Iterative Detection}
We divided the augmented data set into training, validation, and test sets in an 8: 1: 1 ratio, resulting in 80, 10, and 10 UDG candidates, respectively. The training set was utilized to train the model over 200 epochs. {As mentioned above, our UDGnet framework incorporates a data augmentation module. Except for the mosaic augmentation, the data augmentation strategies within the model are consistent with those applied during sample expansion. Therefore, during the initial training with the dataset described in Section \ref{subsec:Sample Expansion}, we set the in-model data augmentation rate to 10\%. In subsequent iterative training stages, this rate was increased to 50\%. It is important to note that the mosaic augmentation rate was consistently maintained at 10\%.}

After 200 training epochs the updated UDGnet-K model is considered fully trained, as shown by the loss curve in Fig. \ref{loss}. Using this model with a confidence threshold sets to 0.5, we have identified a total of 19 UDG candidates within the 10 cutouts comprising the test set. Only 6 of the candidates correspond to the previously labeled UDG test samples, resulting in a recall rate of 60\%.
The visual inspection of the remaining candidates reveals that most of them possess a diffuse surface brightness distribution, typical of UDGs appearance.
Four of these ``newly discovered'' UDG candidates are shown in Fig. \ref{fig6}. However, some others of them also show a rather irregular shape, including the presence of pseudo-arms, incompatible with being canonical UDGs. We have derived the $\mu_e$ and $R_e$ of all these new detections and they all show values within the UDG definition, which suggests that the structural parameters are not enough to define a UDG. Instead, only visual inspection can ensure the genuinity of the new UDG detections and the purity of the extended training set. Hence, in the following, we will primarily use visual inspection to collect the training set, while we will leave the structural parameters as the last selection criteria to adopt to refine the final UDG catalogs.

Going back to the training process, if, on the one hand, the discovery of new UDG candidates on top of the pre-selected test sample suggests that the model has learned the features of UDGs already at this step, on the other hand, the low recovery rate (60\%) also suggests that the model has margins to be improved. 
To do that, we have implemented an iterative detection method, consisting of the following steps:

(1) \textbf{Image grouping:} it consists in splitting the KiDS image catalog of 592,620 chunks (see \S\ref{sec:data}) into 4 groups of 150k, 150k, 150k and 142,62k chunks each;

(2) \textbf{Selection and Detection:} this step consists of applying the UDGnet-K model trained on the initial 100 UDG samples, to detect UDG candidates in the first group of 150k images.

(3) \textbf{Visual Inspection and Labeling:} This consists of visual inspection of newly detected UDG candidates, to add those that are correctly identified back to the initial set of augmented 100 UDG training and to obtain an expanded set of labeled training.

(4) \textbf{Retraining:} here we retrain the UDGnet-K model with the updated training set.

(5) \textbf{Repeating:} in this step we use the retrained UDGnet-K model to detect UDG candidates in the next group of images, and repeat the process from step (2) until the detection has been concluded in the last group of images.

This iterative detection method not only maximizes the utility of existing data but also facilitates the discovery of new UDG candidates, thereby providing a more comprehensive and diverse training set to enhance model training. After the full round of train/detection, we have collected a total of 493 UDG candidates (excluding the data-augmented images).

\subsection{Final Sample Selection} \label{subsec:Selection of Final Sample} 
The expanded training set derived in the previous section is used to retrain the UDGnet-K model a last time to be finally ran to perform a final discovery run across all 592,620 chuncks.

\begin{figure}[]
    \hspace{-0.5cm}
    \includegraphics[width=0.5\textwidth]{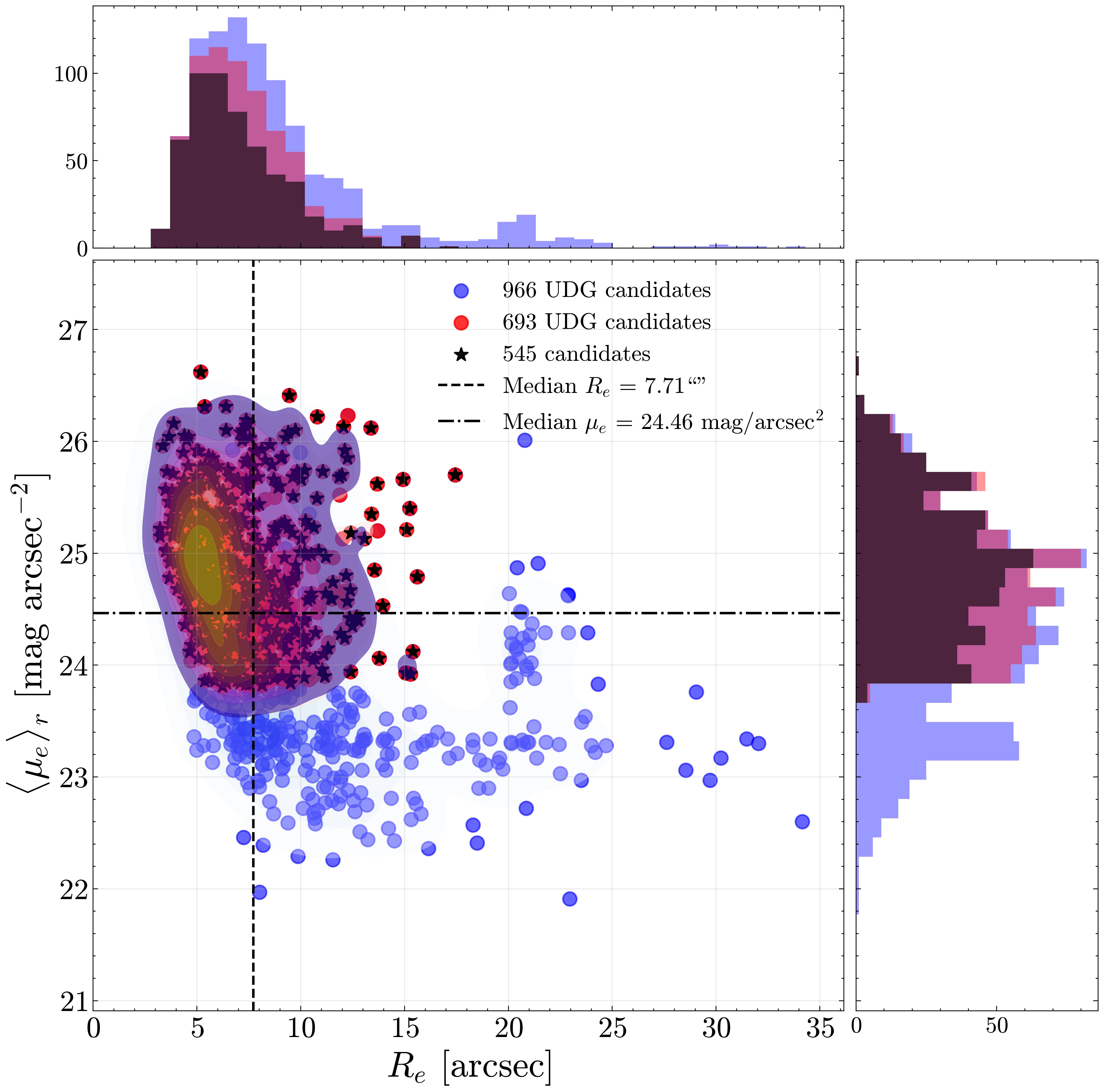}   
    \caption{The distribution of effective radius $R_e$ and mean effective surface brightness $\langle \mu_{e}\rangle_r$ of UDG candidates. The 693 candidates represent candidates which meet the selection criteria $3''<R_e<20''$ and $\langle \mu_{e}\rangle_r>23.8$ mag/arcsec$^2$. The 545 candidates represent the candidates with grad A (see \S\ref{sec:grading}). The black dashed line represents the median of all candidates' $R_e$ and $\langle \mu_{e}\rangle_r$.}
    \label{fig13}
\end{figure}

\begin{figure}
    \hspace{-0.3cm}
    \includegraphics[width=0.5\textwidth]{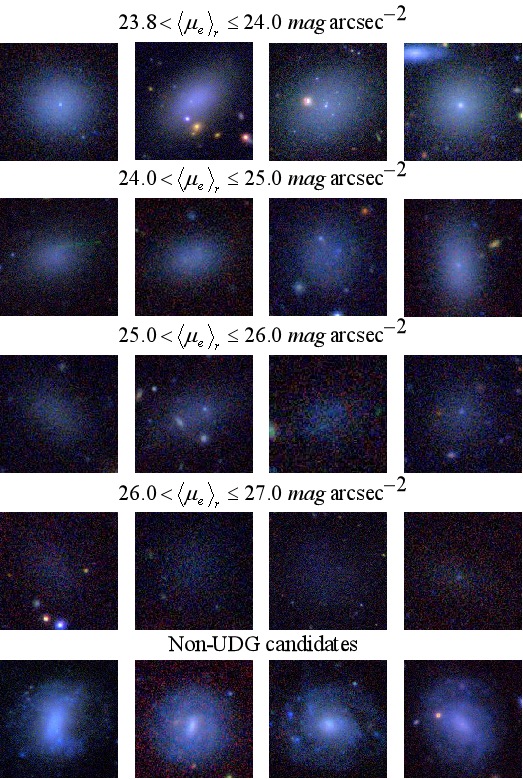}
    \caption{Examples of candidates in different mean effective surface brightness ranges. Each row's mean effective surface brightness falls within a specific range. In the last row, we show some non-UDG candidates. The size of each image is 30$'' \times 30''$.}
    \label{fig8}
\end{figure}

The fully trained UDGnet-K model detected 3,315 potential UDG candidates, although a quick check has shown the presence of bad images, some objects at the edge of the images, and duplicated candidates.
After a cleaning check performed by two of us to delete all critical situations and outlier candidates (including irregular and LSB pseuso-spiral systems), and having also eliminated the duplicates found in the overlapping areas of contiguous chunks, we are left with a total of 966 candidates. By imposing the final criterion to their effective radius ($3''<R_e<20''$) and mean effective surface brightness ($\langle \mu_{e}\rangle_r>23.8$ mag/arcsec$^2$), we are finally left with 693 candidates. In Fig. \ref{fig13} we show the distribution of the $R_e$ and $\mu_e$, as compared to the 966 sample. {We find that the majority of candidates exhibit effective radii concentrated in the range of 6$''$ to 10$''$, consistent with the size distribution of UDG populations reported in other large-scale detection efforts \citep[e.g.,][]{zaritsky2021systematically}. However, we also see a tail of objects with $\langle \mu_{e}\rangle_r>23.8$ mag/arcsec$^2$ and $10''<R_e<15''$ and only a few with $R_e>15''$, while there is a group of systems with $R_e>20''$}. A visual inspection of these systems has shown that the majority of them have a clear spiral-like structure, while there are four of them that still look like UDGs, with quite resolved stars in them, suggesting them to be very nearby systems, which, assuming 1.5 kpc as the lower limit for UDGs, and standard Planck $\Lambda$CDM cosmology \citep{Planck2020}, this implies a lower redshift limit of would translate in a scale of 1.5kpc/21$''\sim 0.075$ kpc/arcsec, or a lower distance bound of 15 Mpc, almost equivalent to the Virgo Cluster. The median value of $R_e$ for our candidates is 7.7$''$, which, assuming, again, 1.5 kpc as the lower limit for UDGs, corresponds to a physical scale of $1.5 / 7.7 = 0.195$ kpc arcsec$^{-1}$. Under a standard Planck $\Lambda$CDM cosmology \citep{Planck2020}, this implies a lower redshift limit of $z \sim 0.01$ (approximately 43.5 Mpc), suggesting that a large fraction of our UDG candidates are likely located in the nearby Universe (see \S\ref{sec:matching} and \ref{sec:spatial_distribution}).
A random selection of the 693 UDG candidates is shown in Fig. \ref{fig8}, where they are separated according to their $\langle \mu_{e}\rangle_r$. Among these we still can see some dubious objects with irregular features, and residual of spiral arms. To select an High-Quality (HQ) sample with minimal contamination from non-UDGs we have performed a full visual review of the 693 objects.

\begin{table}[]
\centering
    \footnotesize
    \setlength{\tabcolsep}{1.6pt}
    \caption{The catalog of some high-quality UDG candidates.}
\begin{tabular}{ccccccc}
    \hline\hline
    R.A.         & Dec.   & $ \langle \mu_{e}\rangle_{r}$ & $R_{e}(r)$ & mean\_score & $ \sigma $ & $\langle {\rm Grade} \rangle$ \\ 
    deg  & deg  & $\ mag\ arcsec^{-2}$ & $arcsec$ & & &  \\
    \hline
    0.627707  & -30.6654 & 24.32   & 5.55       & 7.4         & 2.88       & A \\
    0.726513  & -33.999  & 25.01   & 5.52       & 7.71        & 2.56       & A \\
    1.899134  & -32.168  & 24.87   & 4.2        & 8.38        & 2.56       & A \\
    2.142476  & -33.8264 & 25.02   & 3.49       & 8.71        & 1.6        & A \\
    2.162882  & -33.8838 & 25.00   & 5.49       & 9.57        & 1.13       & A \\
    7.761782  & -33.2707 & 23.93   & 15.06      & 10.0        & 0          & A \\
    7.980486  & -33.2661 & 24.43   & 13.95      & 10.0        & 0          & A \\
    8.242036  & -32.764  & 25.38   & 9.22       & 9.14        & 1.46       & A \\
    8.404569  & -32.3808 & 24.12   & 6.29       & 9.4         & 1.34       & A \\
    8.779283  & -28.0503 & 25.66   & 14.93      & 10.0        & 0          & A \\
    10.549599 & -31.6502 & 24.60   & 11.44      & 7.4         & 2.88       & A \\
    11.680739 & -31.5162 & 25.39   & 6.33       & 9.57        & 1.13       & A \\
    14.161699 & -31.786  & 24.68   & 10.33      & 8.8         & 1.64       & A \\
    15.9952   & -28.5139 & 24.28   & 6.66       & 8.2         & 1.64       & A \\
    18.562234 & -32.4841 & 24.63   & 4.84       & 9.4         & 1.34       & A \\
    18.603041 & -32.6282 & 24.82   & 5.72       & 9.4         & 1.34       & A \\
    18.639711 & -32.1925 & 24.99   & 4.48       & 7.14        & 3.13       & A \\
    18.671723 & -32.2038 & 24.26   & 9.25       & 10.0        & 0          & A \\
    19.011245 & -31.7601 & 24.57   & 4.56       & 8.12        & 1.55       & A \\
    19.213181 & -33.0282 & 24.59   & 4.39       & 9.4         & 1.34       & A \\
    19.496324 & -32.769  & 24.61   & 5.57       & 9.14        & 1.46       & A \\
    19.830581 & -33.0867 & 25.01   & 5.0         & 9.14       & 1.46       & A \\
    20.117843 & -31.4713 & 25.92   & 7.83       & 9.57        & 1.13       & A \\
    20.266866 & -33.157 & 24.25   & 8.7       & 8.5        & 1.73       & A \\
   ... & ... & ...   & ...       & ...       & ...     & ... \\
        
    \hline
\end{tabular}
\label{tab:table1}
\end{table}

\begin{figure}[]
    \centering
    \includegraphics[width=0.48\textwidth]{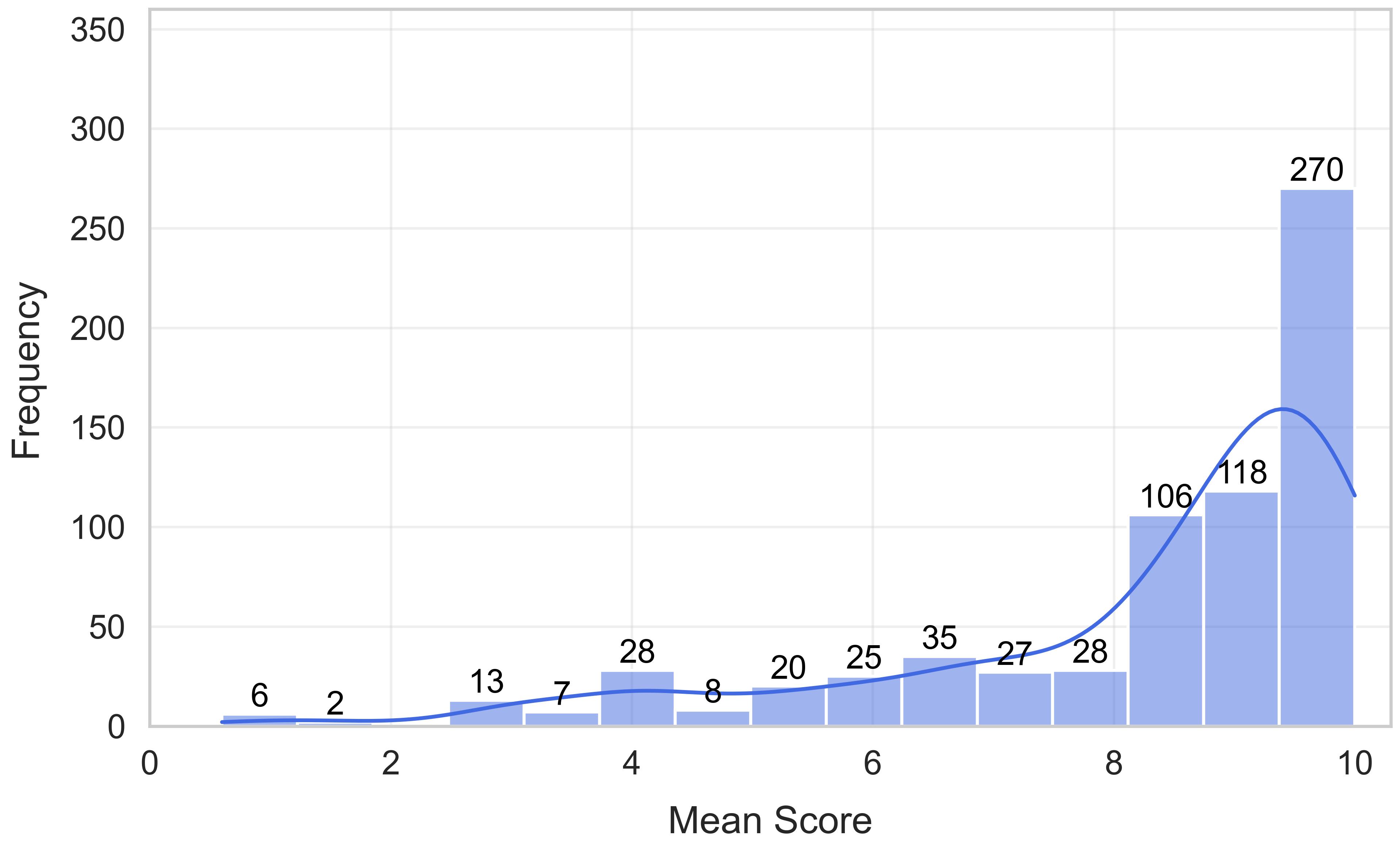}
    \caption{Score distribution of UDG candidates.The histogram shows the number of sources at different score levels, while the blue curve represents the trend of the distribution.}
    \label{fig12}
\end{figure}

\begin{figure*}
    \centering
    \includegraphics[width=0.99\textwidth]{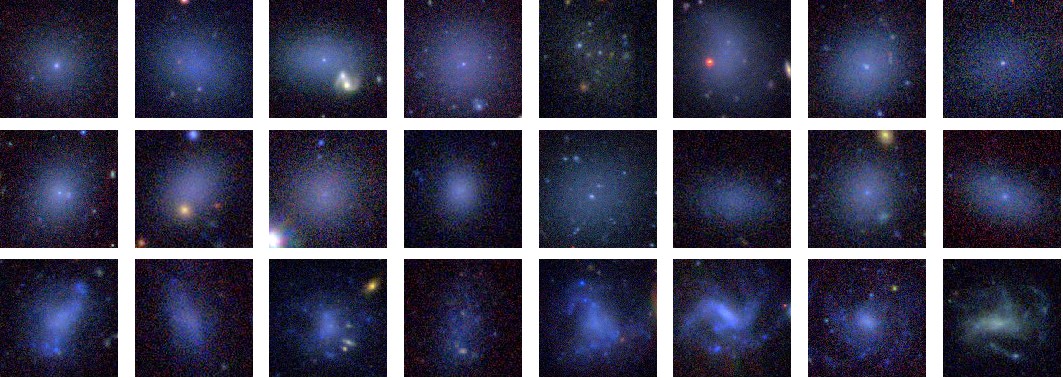}
    \caption{The representative examples of UDG candidates by category (A, B, and C). In the first two rows we show 16 UDG candidates of category A, and in the last row we show 4 candidates of category B and 4 candidates of category C.}
    \label{fig12b}
\end{figure*}

\section{High-quality sample} \label{sec:Final Sample Selection}
In this section, we report the results of the grading of the 693 candidates. We remind that this sample is obtained from the original ``potential'' UDG candidates (3315) from the direct application of the UDGnet-K, from which we have derived a ``cleaned'' sample of 966 objects that reduced to 693 by imposing the adopted constraints on $R_e$ and $\langle \mu_{e}\rangle_r$. The ranking of this latter sample will define the HQ sample for which we study the spatial distribution across the KiDS area and drive some conclusions about their tentative environment.

\subsection{Visual grading} \label{sec:grading}

The UDG candidate sample was graded by eight inspectors, who evaluated the color-combined images of the candidates. Each candidate was assigned to one of four grades, corresponding to scores indicated in brackets: {\it a}. secure UDG (10), {\it b}. probably an UDG  (7), {\it c}. probably not an UDG (3), {\it d}. non-UDG (0). 

The overall morphological criteria are the regularity of the surface brightness and the absence of major substructures in it. Other criteria to define the quality are the presence of bright sources or the background level/structure, that might reduce the confidence on the classification. We stress here that these visual criteria might represent a biased definition of UDGs, that should principally consist on structural criteria based on size and surface brightness. Indeed, as stressed by \citet{prole2019observational}, field UDGs defined in their sample show more irregular shapes and signs of star formation, e.g. on the form of blue knots (as well as bluer colors). In this respect, with our visual inspection we might be introducing some selection bias, if these ``irregular'' systems are indeed genuine UDGs. We keep this epistemological note in mind as, in absence of a defined formation scenario of these objects, it is hard to assess whether our conservative approach to retain UDGs as regular systems without substructure is really a different class with respect to more generic ``irregular LSB'' systems. We will return to this matter later. 

The final score of each UDG candidate is obtained by averaging the score of the 8 inspectors. These average values have been used to define an ``average'' grade ($\langle {\rm Grade} \rangle$) defined as: C for average scores of (0-3), B for (3-7), and A for (7-10). Hence, a $\langle {\rm Grade} \rangle=$A object is an object where all inspectors have given {\it b}-grade or higher, roughly speaking, or 6/8 have $a$ or $b$ grade with majority of $a$. In Table \ref{tab:table1} we show a subsample of the final catalog of 693 candidates, where we report the main information including the coordinates (RA and Dec), the mean effective surface brightness $\langle \mu_{e}\rangle_r$, the effective radius $ R_e$, the average score and the related standard deviation $\sigma $ and $\langle {\rm Grade} \rangle$. 

In this catalog, the number of UDG candidates in categories A, B, and C are 545, 127, and 21, respectively. Interestingly, in our sample, there is a residual minority of non-UDGs, meaning that the ``cleaning'' step effectively eliminated most of the spurious detections. 

\begin{figure}[]
    \centering
    \includegraphics[width=0.48\textwidth]{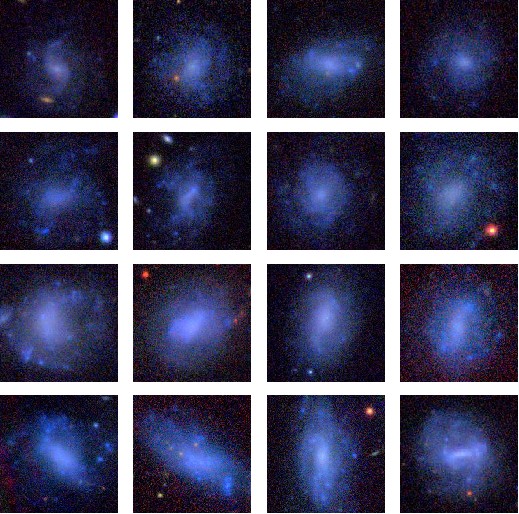}
    \caption{UDG classification consistency and surface brightness properties. The top two rows show increased disagreement ($\sigma>2$) for lower-scoring candidates (mean score $<7$). The bottom 2 rows show the candidates with low-scoring and high surface brightness ($\langle \mu_{e}\rangle_r<24 \rm ~mag\rm ~arcsec^{-2}$).}
    \label{high_sigma}
\end{figure}

\begin{figure}[]
    \includegraphics[width=0.48\textwidth]{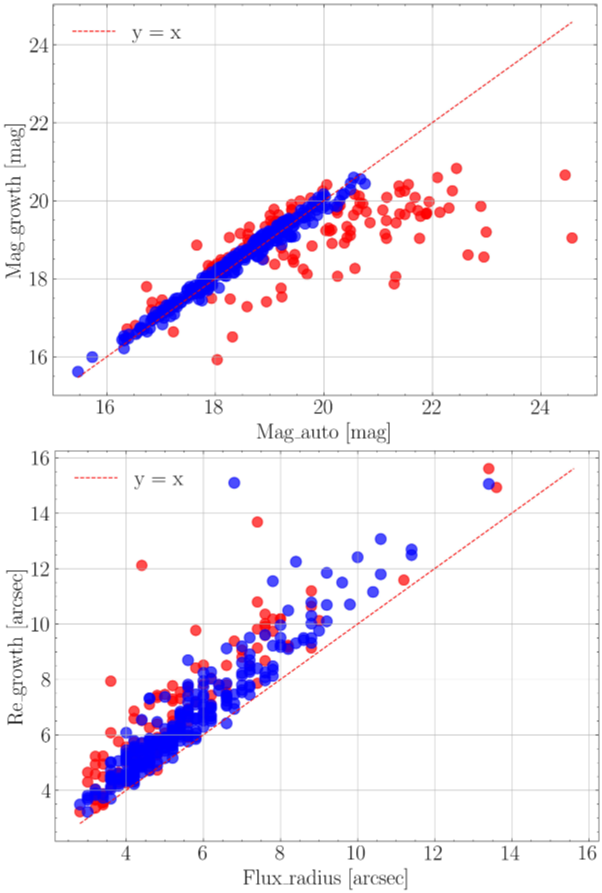}   
    \caption{Comparison of structural parameters (Mag and $R_{ e}$) between the HQ UDG candidates and their counterparts in the KiDS DR5 catalog. {\it Top:} The total $r$-band magnitudes derived from our growth curve method versus the KiDS catalog {\tt mag\_auto} values. {\it Bottom:}  $R_{e}$ estimated from our growth curve compared with the KiDS catalog {\tt flux\_radius}. Red points indicate systems with large photometric uncertainties ($>0.33$ mag).}
    \label{match_with_kids}
\end{figure}

The score distribution of the candidates is shown in Fig. \ref{fig12}, where we can see that more than half of the candidates, 270+118=388, have been judged as a sure UDG from almost all inspectors (average score larger than 9), while 28+106=134, have been judged to be sure UDG from about half of the inspectors (average score between 8 and 9). Overall, the A sample of 545 candidates can be considered a {\it golden} sample of HQ candidates. A gallery of these HQ UDGs is shown in Fig. \ref{fig12b}, where we also add 4 B and 4 C graded systems for comparison.

Looking into the details of the scoring, we find some correlations, partially expected. First, the lower the score, the larger the scatter among the inspectors, meaning that the visual definition of the UDG becomes more debated for unclear cases. For instance, almost all mean scores lower than 7 have a scatter larger than 2. By re-checking some of these objects, they all show substructures in their surface brightness distribution, like knots or signatures of spiral arms (see Fig. \ref{high_sigma}). Although the presence of knots might still be consistent with the presence of star-forming regions mentioned earlier in this paper and in literature for field systems (see again \citealt{prole2019observational}), the presence of spiral-like structures or streams suggests that these candidates are reasonably to be discarded from the UDG sample. Hence, to conclude our epistemological note, we decide to maintain the conservative approach of excluding clearly irregular systems from our golden sample, accepting that this could reduce the completeness of our sample compared to other less conservative collections, but aiming at optimizing the purity. For example, \citet{prole2019observational} claim the detection of $\sim200$ UDGs in 39 deg$^2$ (or $\sim$5 UDG per deg$^2$), which is almost a factor of 10 larger than our $\sim550$ UDGs in 1350 deg$^2$ (i.e. $\sim$0.4 UDG per deg$^2$). This factor is too large to be real, especially considering that the Prole et al. work refers to ``field'' candidates. We believe that this can be a consequence of the absence of any visual criteria to define an UDG sample as, by looking at the example of UDG candidates in their paper. Also, less than half of their candidates would be qualified as HQ systems in our visual grading. To corroborate this conclusion, we notice that a sample of UDG candidates in low-to-moderate density fields from \citet{2021MATLAS_UDG}, have found, after a visual classification, 0.4 UDG/deg$^2$. 

We finally find that there is no correlation of the score with $\langle \mu_{e}\rangle_r$, except that the larger fraction of the low scores ($<7$) reside in the brightest $\langle \mu_{e}\rangle_r$ bin. Some of these objects are also shown in Fig. \ref{high_sigma}, where we can see that, even in this case, the objects show substructures and spiral-like structures that have driven the low grading.

\subsection{Matching with the KiDS catalog} 
\label{sec:matching_kids}
UDGs are objects that have intrinsically low signal-to-noise ratio (SNR), due to their diffuse light distribution. However, 
regardless of their intrinsic luminosity, they should be eventually detected in KiDS images from the detection pipeline (see e.g. \citealt{de2015first}), with only the lowest SB ones (see e.g. Fig. \ref{fig8}) being missed by the KiDS sourcelists. In order to check this for the HQ sample, we have matched their coordinates with the KiDS-DR5 catalog (\citealt{wright2024fifth}). We have used a matching radius of 2$''$, which is big enough to account for centroid errors, and small enough to minimize the mismatch with other background/foreground projected sources which can be close enough to the UDG centers (see e.g. Figs. \ref{fig11} and \ref{fig8}). With this matching radius, we have obtained 385/545 matches, i.e. 70\% of the HQ sample. However, using a less conservative matching radius of 3$''$, we have matched 465/545 HQ candidates, corresponding to about 85\% of the full sample. Very likely a small fraction of these latter might indeed contain some mismatches, hence, for the following analysis, we will consider the 2$''$ matched sample. In both case, it seems clear that the UDG detection is possible even for standard general-purpose source extraction algorithms with no specific optimization for faint systems, but at the cost of a rather heavy incompleteness and contamination (see below). Indeed, the selection of UDG candidates in these catalogs should be based on the availability of reliable structural parameters that can be used to check the UDG criteria based on their size and surface brightness. For instance, in KiDS, besides multi-band aperture and Kron photometry, no effective radii is provided in the released catalogs. Suitable parameters to identify UDGs from the KiDS catalogs 
are the {\tt mag\_auto} and the {\tt flux\_radius} from Sextractor (\citealt{bertin1996sextractor}) as a proxy of the total $r$-band magnitude and the circularised half-light radius, respectively. In the top panel of Fig. \ref{match_with_kids} we show that, if we consider the sources with mean SNR$>9$, corresponding to  {\tt mag\_auto} errors of the order of 0.33 mag, the {\tt mag\_auto} is well consistent with our total $r$-band magnitude estimate from the growth curve, with a 
mean difference of $0.12\pm0.24$ mag, i.e. slightly overestimating the magnitude although consistent with zero difference within 1$\sigma$. Below the SNR=9, the {\tt mag\_auto} estimates start to become more scattered, as shown in the bottom panel of Fig. \ref{match_with_kids} (red dots). 
In the same figure, we also show that the {\tt flux\_radius} parameter is a biased proxy of the effective radius, here represented by our estimate from the growth curve. In particular, it is evident a $\sim20\%$ underestimate of the Sextractor parameter, consistently with earlier finding of strong systematics affecting this parameter for LSB systems (e.g. \citealt{2025_Thuruthipilly_fluxrad}).

Therefore, blind searches based only on the use of standard catalog structural parameters are unreliable to select clean samples of UDG candidates. Indeed, 
the underestimate of the effective radii would cause brightest $\langle \mu_{e}\rangle_r$ producing a loss of candidates. Also, should one rely on photometric parameters, a safe cut on SNR would be needed to limit the contamination from noisy parameters. In this case, we have seen this would also produce the loss of relevant low SNR candidates. Hence, the use of deep learning remains a viable and accurate alternative to the human visual search, to select complete samples of UDG candidates over large areas of the sky. 

A final thing to mention are photometric redshifts. The KiDS DR5 catalog, as all previous releases, provide photometric redshifts from 9-band Gaussianized apertures, from the Bayesian photo-z software \citep[BPZ,][]{2000_BPZ}). These represent a crucial ingredient for weak lensing studies, where galaxies are usually grouped in rather large redshift bins, that exceed the typical redshift errors $\sigma_{\Delta z}=0.05$ (see \citealt{wright2024fifth}). In principle, we could use the same photo-z to associate a tentative distance to the matched UDG candidates with KiDS, as done for {\tt mag\_auto} and {\tt flux\_radius}. However, these errors are overwhelmingly higher that the true intrinsic redshift of these systems, especially if located at low redshifts. To give a perspective, an UDG at $z\sim0.05$ (corresponding to $\sim$ 220 Mpc) would have a 100\% error in redshifts, which more or less corresponds to a 100\% error on the effective radius in kpc, making the use of photo$-z$ basically pointless. For this reason, we have decided to exclude the use of published photo$-z$ in this analysis, hereafter.  

\begin{table*}[]
\centering
    \setlength{\tabcolsep}{8pt}
    \caption{Cross-match Results of High-Quality UDG Candidates with nearby clusters/groups.}
\begin{tabular}{lcccccc}
    \hline\hline
    Cluster/Group & RA & Dec & Redshift & Dist (Mpc)& $N_{\rm UDG}$ (new) & $N_{\rm UDG}$ matched/total (reference)    \\ 
    \hline
    \multicolumn{7}{c}{With previous literature}\\
    \hline

    Fornax     & 54.621  & -35.450 & 0.0062   &26.8 & 3 & 1+1/12+1 \citep{venhola2017fornax}$^*$     \\
    NGC 4690   & 191.981 & -1.656  & 0.0092   &40.1 & 2  & 2/2 \citep{marleau2021ultra}    \\
    NGC 5846   & 226.622 & 1.605   & 0.0062   &26.8 & 2  & 3/7 \citep{marleau2021ultra}$^{**}$   \\
    NGC 5576   & 215.265 & 3.271   & 0.0051   &22.5 & 2  & 2/2 \citep{marleau2021ultra}    \\
    \hline
    \multicolumn{7}{c}{Without previous literature}\\
    \hline

    IC 1860   & 42.391  & -31.189  & 0.022     &96.5 & 2  &                                \\
    IC 5157   & 330.862  & -34.941 & 0.0147    &64.1 & 4  &                                \\
    IC 5358   & 356.938  & -28.141 & 0.028     &123.0 & 2  &                                \\
    NGC 0439   & 18.447  & -31.747 & 0.019     &83.1 &5 &                                 \\
    NGC 4073   & 181.112 & 1.895   & 0.020     &87.5 & 15 &                                 \\
    NGC 4636   & 190.707 & 2.687   & 0.004     &17.5 & 5  &                                 \\
    NGC 7176   & 330.535 & -31.990 & 0.0084    &36.4 & 6 &                                \\
    NGC 7507   & 348.031 & -28.539 & 0.0053    &22.9 & 2  &                                \\
    UGC 07813   & 189.755 & 0.366  & 0.023     &100.8 & 4  &                                \\
    ESO466-021 & 329.479 & -28.805 & 0.023     &100.8 & 1  &                                \\
    2MASS J2259-3334 & 344.755 & -33.572 & 0.028 &123.0   & 3  &                                \\
    
    \hline
    \label{tabel2}
\end{tabular}

\begin{flushleft}
* The catalog from \citet{prole2019halo}  provided the same match as the \cite{venhola2017fornax}. paper. The +1 candidate is in the Venhola catalog but classified as normal LSBG (see text for more details). \\

** One of the MATLAS UDGs was earlier discovered by \cite{forbes2019ultra}.

\end{flushleft}

\end{table*}

\subsection{Matching with external catalogs from literature} 
\label{sec:matching}
As a further validation of our catalog, in this section we report the results of the match of the HQ candidates listed in Table 1, with some external catalogs reported in the literature, which overlap with the KiDS footprint. In particular, we use a local UDG samples from \cite{venhola2017fornax}, \cite{prole2019halo} and \cite{marleau2021ultra}, which report candidates belonging to low redshift groups/clusters. The prominence of local candidates is not surprising as, due to the LSB nature of these systems, and intrinsic faintness, distance is a natural selection effect as for elusive systems, angular size helps the identification of convincing candidates, both for automatic searches (e.g. setting the number of pixel with a sufficient SNR as a criterion for a detection) and for visual inspections (where small LSB systems can be confused with correlated noise). We have also seen that more than half of our candidates have angular sizes that might be compatible with low redshifts (see \S\ref{sec:struc_par}), thus making the chance to select nearby systems higher.  

\citet[][V+17 hereafter]{venhola2017fornax} has collected a catalog of $\sim 200$ LSBGs including 12 UDGs, which we have cross-matched with our HQ catalog. We have found 2 matches, of which only one candidate is classified as UDG in their catalog (FDS10\_LSB52), while the other one (FDS11\_LSB49) is slightly off the UDG range being $R_e<1.5$ kpc within the errors. Indeed, a small scatter of the measurement around the 1.5 kpc threshold can make a standard LSBG to jump into the UDG sample, meaning that the classification of an object can change from dataset to dataset.

Assuming the same distance of Fornax of V+17, our matched systems all are confirmed to be UDGs with the $R_e=1.86,~1.62$ kpc and $\langle \mu_{e}\rangle_r=25.21,~24.35$ mag/arcsec$^2$, respectively.  
All the others V+17 candidates cannot be matched because they are distributed in the $\sim$3/4 of the cluster area, which is outside of the KiDS area. 
Interestingly, looking at the position on the sky of our HQ candidates (see also next Section), there are more UDG candidates in our catalog that might be associated to this cluster. Following V+17 footprint, we have checked which ones of the UDG candidates within 3 deg from the Fornax Brightest Cluster Galaxy (BCG, NGC 1399) possess a $R_{e}>1.4$ kpc assuming the same distance of the BCG. We have used 1.4 kpc to account for the possibility that the UDG has a variation of the distance along the line-of-sight within the cluster. We have found that 5 candidates (including the one already matched) are compatible with being  UDG candidates of the Fornax cluster, 3 of which are new candidates missed in previous collections. All above information is reported in Table \ref{tabel2}.

The cross-match with the MATLAS UDG catalog from \citet{marleau2021ultra}, making use of HST imaging, has returned a total of seven matched candidates. Among them, two UDG candidates are matched with the two UDGs belonging to NGC 4690 in their catalog, two with the two associated to NGC 5576, meaning that we have recovered the known samples around these two galaxies.
We have also matched three of our UDG candidates with the catalog of five candidates associated to NGC 5846. In this case, the HQ catalog misses four objects that, at the visual inspection, are too faint in KiDS with respect to the corresponding HST images.  
As for Fornax cluster, we have checked if other UDGs from our catalog are compatible with being associated to the same groups. In this case we have used the galaxy distances to define a radius of 1 Mpc to select UDG candidates and convert the $R_e$ to kpc, using again 1.4 kpc as a lower limit to define the candidates as UDGs belonging to the group. As we have reported in Table \ref{tabel2}, we have found that three new UDGs are found in NGC 5576, three in NGC 4690 and two in NGC 5846. The coordinates and main properties of these new UDG members are reported in Appendix \ref{apppendix}. 

\begin{figure*}[ht!]
    \centering
    \includegraphics[width=1.02\textwidth]{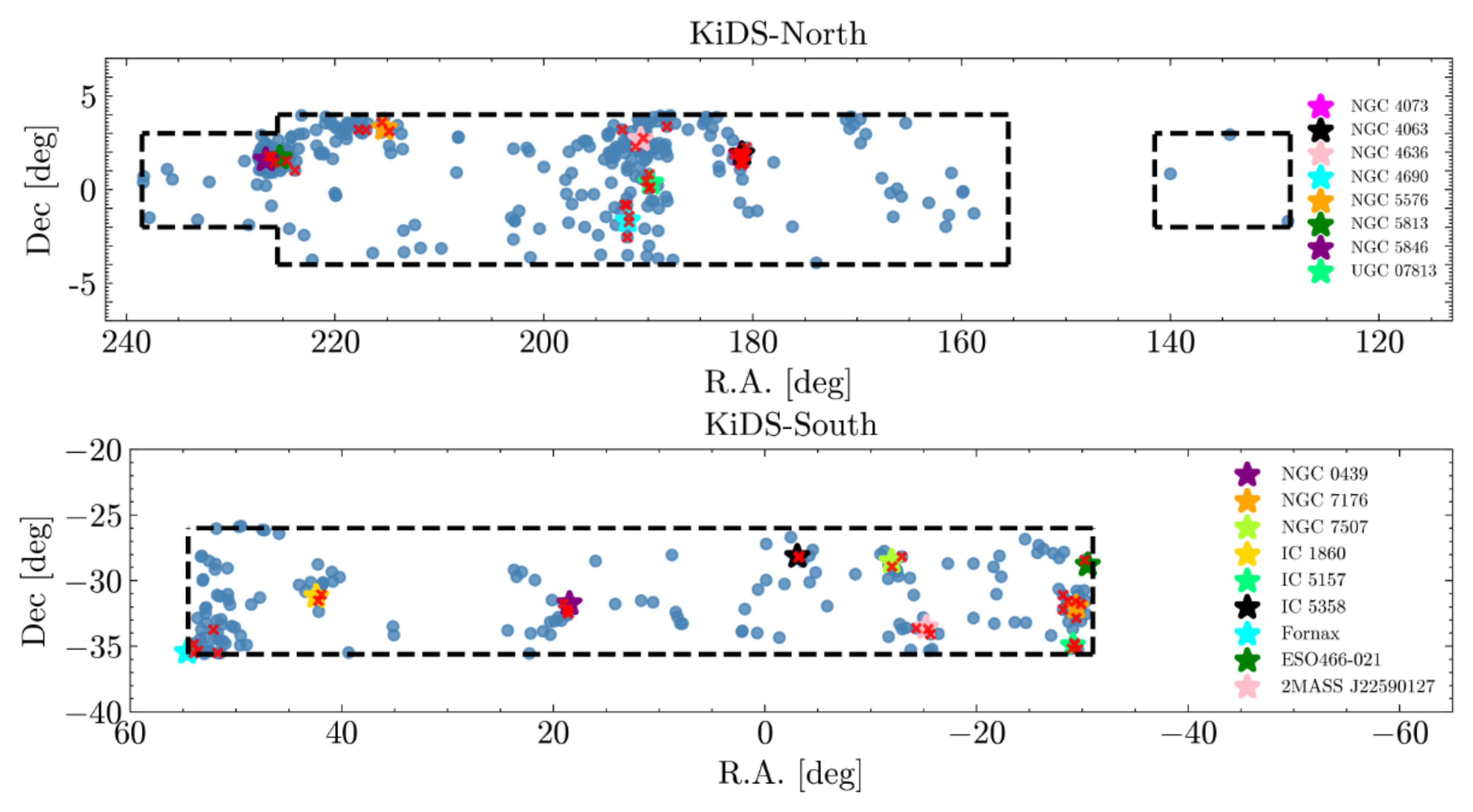}
    \caption{The distribution of A-grade UDG candidates in KiDS DR5. The red crosses denote UDG candidates matched to known galaxy clusters.}
    \label{all_cluster}
\end{figure*} 

To conclude this section, we can use the matched sample of nine known UDGs (last column on the right in Table \ref{tabel2}) for a sanity check of our structural parameters, by comparing the inferences of $R_e$ and absolute magnitude (or $L_{\rm max}$) from Eq. \ref{eq2} with the values reported in previous litarature. We find a $\delta R_e=R_{\rm HQ}-R_{\rm lit}=-0.09\pm0.11$, where $R_{\rm HQ}$ is the effective radius from our catalog converted into kpc, assuming the associated group distance, and $R_{\rm lit}$ is the same radius from literature, and a $\delta m_r=m_{\rm HQ}-m_{\rm lit}=0.06\pm0.19$, where $m_{\rm HQ}$ is the $r$-band total magnitude from our catalog and the $m_{\rm lit}$ is the same from literature. Interestingly, both quantities show no systematic deviations within the statistical errors, even though our parameters are based on a less sophisticated procedure than the 2D surface fitting adopted in literature. 

\subsection{Spatial distribution and environment and associations in the local universe} \label{sec:spatial_distribution}
In the previous section, we have seen that our HQ catalog contains matches with previously detected UDG candidates in Fornax cluster, NGC4690, NGC 5846, and NGC 5576 groups. The KiDS footprint covers, indeed, a large variety of environments, including many other close groups and clusters with which we might expect the UDG candidates to be associated. The distribution of the HQ candidates is illustrated in Fig. \ref{all_cluster}.

From this figure, we can see that the candidates are not uniformly distributed but seem to cluster in specific regions, with locally higher number densities, that generally fade into lower density regions. This suggests that the majority of UDGs are associated to overdensities tracing some large scale structures. The great advantage of covering large areas of the sky is that KiDS can provide also many ``field'' UDGs, as pointed out for a general population of dwarf galaxies in \cite{marleau2025euclid}. Despite the great scientific value of the latter ones, the UDGs inhabiting the high-density regions are more likely associated to nearby systems for which the distance is known.   
To check for these possible associations, we have followed the same approach as in the previous section for the known UDGs, but using the catalog of bright nearby galaxies (i.e. absolute B-magnitude $M_B<-20$\footnote{From Hyperleda: https://leda.univ-lyon1.fr/}). For each of these galaxies, we have searched for the UDGs within a projected radius of 1 Mpc having a $R_e>1.4\rm~ kpc$, assuming for them the same distance as the large host galaxy. The results are shown in Table \ref{tabel2}, where we report the ID of galaxies and clusters being compatible with hosting UDGs and the number of them. These are also plotted in Fig. \ref{all_cluster} where we see the nearby galaxies placing themselves in the center of the UDG overdensities, confirming the impression that a large fraction of UDGs have to be satellites of galaxy/group/cluster halos. In this work, we are not interested in a complete characterization of the environment of the UDG candidates, which we leave for a forthcoming paper. However, using the lesson learned in the previous section, it is important to assign an environment and a distance to a number of candidates in our catalog to have a confirmation of the nature of UDGs of these systems. As seen in Table \ref{tabel2}, we could associate only 67 UDG candidates because we have limited our choice to nearby systems ($v_{\rm rad}<8000$ km/s), which matched almost all the high-density patches in Fig. \ref{all_cluster}. This number could be larger if we had considered a larger area around them (i.e. larger than 1 Mpc), or if we checked further away groups/clusters. As a first attempt to prelude a deeper and more attentive analysis of the environment for all the UDGs in the catalog, we just stress that we have an interesting first sample of UDG members in 15 nearby groups/clusters. Among these, in the KiDS-North we remark a large concentration around the area connecting the NGC 5846 group and NGC 5576 group, that is worth to be investigated in more details, and the southern tails of the Virgo cluster (represented by NGC 4636), with an interesting filament extending toward NGC 4063.
In the KiDS-South it is evident the quarter of sphere around the center of the Fornax cluster, and other larger concentrations around NGC 7176, NGC 7507 and 2MASS-J22590127.

The details of all the associated UDG candidates is again reported in the Appendix \ref{apppendix}, together with the ones that have been found in the previous section.

\section{Summary} \label{sec:Summary}
In this paper, we have presented UDGnet, a new deep learning framework tailored for the large-scale detection of UDGs in photometric imaging data. This framework builds upon the previously proposed LSBGnet framework, which was originally developed for the identification of LSBGs. Based on UDGnet, we construct a specialized model, UDGnet-K, specialized for application to the KiDS DR5 dataset. The direct application of the UDGnet-K to the full KiDS footprint, consisting on 1350 deg$^2$, has produced a sample of 3,315 UDG candidates. After removal of duplicate and spurious detections, we have obtained a final sample of 966 objects. For all candidates we have measured 
the effective radius, $R_e$, and the mean effective surface brightness, $\langle \mu_e\rangle_r$, based on the luminosity growth curve in the images of the $r-$ band, and selected UDG candidates based on reasonable selection criteria. The derived parameters exhibit no systematic deviations within the statistical uncertainties.

We have opted for selection threshold to $\langle \mu_e\rangle_r\geq 23.8~\rm ~mag\rm arcsec^{-2} $ and $ 3''\leq R_{e}(r)\leq 20''$. This brought us a catalog of 693 UDG candidates visually inspected by 8 evaluators. We have finally obtained high-quality 545 UDG candidates, corresponding to a frequency of $\sim0.4$ UDG/deg$^2$, consistent with other findings using visual inspection to obtain clean UDG samples.

The spatial distribution of UDG candidates across the KiDS footprint is rather inhomogeneous, with noticeable over-densities in some regions and others showing very few or no candidates per deg$^2$.  After cross-matching with existing catalogs of known local UDGs, we independently re-identified 9 previously reported UDGs located in the Fornax cluster, as well as in the NGC 5846 and NGC 5576 (see Table \ref{tabel2}). In addition, we identified new UDG candidates within the same clusters/groups. This known sample has proven valuable for validating the robustness of our structural parameter measurements ($R_e$ and $\langle \mu_{e}\rangle_r$), which show excellent agreement—with negligible systematics—when compared to literature values, primarily derived using GALFIT.
 
We have finally searched for more associations with nearby galaxy clusters, based on geometrical arguments, i.e. being within $\sim 1$ Mpc from the group/cluster center and using the cluster redshift to select those objects whose $Re>1.5$ kpc being at that redshift.
This approach led to the identification of $67$ new UDG candidates in several clusters, including the Fornax, ESO466-021, UGC 07813 and NGC groups, with a notably higher spatial number density compared to field regions, suggesting a strong environmental dependence for UDG occurrence. The full HQ catalog and the associations in the local universe are provided in Tables \ref{tab:table1} and \ref{tabel2}, respectively.

The UDGnet-K model we developed inherits the advantages of the LSBGnet framework, which enables automated, end-to-end large-scale detection for LSBGs. 

Using KiDS DR5 as a test-bench, we have fully assessed a very effective 
iterative detection method allowing for large-scale detection of specific objects despite a lack of sufficient known samples, providing an effective approach for subsequent detection of specific objects.
These new UDGnets can be the foundation for models to be developed for upcoming all-sky surveys like Rubin Observatory/LSST, Euclid and the China Space Station telescope (CSST).

\begin{acknowledgements}
      This study was supported by Shandong Province Natural Science Foundation grant Nos. ZR2022MA089. SH acknowledges the support of the China Scholarship Council (grant n. 202406220048).  Rossella Ragusa acknowledges support grants through INAF-WEAVE StePS founds and through PRIN-MIUR 2020SKSTHZ. NRN acknowledges support from acknowledges support from the Guangdong Science Foundation grant (ID: 2022A1515012251). 
\end{acknowledgements}

\bibliography{aanda}{}

\begin{thebibliography}{84}
\expandafter\ifx\csname natexlab\endcsname\relax\def\natexlab#1{#1}\fi

\bibitem[{Abell {et~al.}(2009)Abell, Allison, Anderson, Andrew, Angel, Armus, Arnett, Asztalos, Axelrod, Bailey, {et~al.}}]{abell2009lsst}
Abell, P.~A., Allison, J., Anderson, S.~F., {et~al.} 2009

\bibitem[{Alabi {et~al.}(2020)Alabi, Romanowsky, Forbes, Brodie, \& Okabe}]{alabi2020expanded}
Alabi, A.~B., Romanowsky, A.~J., Forbes, D.~A., Brodie, J.~P., \& Okabe, N. 2020, Monthly Notices of the Royal Astronomical Society, 496, 3182

\bibitem[{Amendola {et~al.}(2018)Amendola, Appleby, Avgoustidis, Bacon, Baker, Baldi, Bartolo, Blanchard, Bonvin, Borgani, {et~al.}}]{amendola2018cosmology}
Amendola, L., Appleby, S., Avgoustidis, A., {et~al.} 2018, Living reviews in relativity, 21, 1

\bibitem[{Amorisco \& Loeb(2016)}]{amorisco2016ultradiffuse}
Amorisco, N.~C. \& Loeb, A. 2016, Monthly Notices of the Royal Astronomical Society: Letters, 459, L51

\bibitem[{Bautista {et~al.}(2023)Bautista, Koda, Yagi, Komiyama, \& Yamanoi}]{bautista2023ultradiffuse}
Bautista, J. M.~G., Koda, J., Yagi, M., Komiyama, Y., \& Yamanoi, H. 2023, The Astrophysical Journal Supplement Series, 267, 10

\bibitem[{Bellazzini {et~al.}(2017)Bellazzini, Belokurov, Magrini, Fraternali, Testa, Beccari, Marchetti, \& Carini}]{bellazzini2017redshift}
Bellazzini, M., Belokurov, V., Magrini, L., {et~al.} 2017, Monthly Notices of the Royal Astronomical Society, 467, 3751

\bibitem[{{Ben{\'\i}tez}(2000)}]{2000_BPZ}
{Ben{\'\i}tez}, N. 2000, \apj, 536, 571

\bibitem[{Bennet {et~al.}(2017)Bennet, Sand, Crnojevi{\'c}, Spekkens, Zaritsky, \& Karunakaran}]{bennet2017discovery}
Bennet, P., Sand, D., Crnojevi{\'c}, D., {et~al.} 2017, The Astrophysical Journal, 850, 109

\bibitem[{Bennet {et~al.}(2018)Bennet, Sand, Zaritsky, Crnojevi{\'c}, Spekkens, \& Karunakaran}]{bennet2018evidence}
Bennet, P., Sand, D.~J., Zaritsky, D., {et~al.} 2018, The Astrophysical Journal Letters, 866, L11

\bibitem[{Bertin \& Arnouts(1996)}]{bertin1996sextractor}
Bertin, E. \& Arnouts, S. 1996, Astronomy and astrophysics supplement series, 117, 393

\bibitem[{Borlaff {et~al.}(2022)Borlaff, G{\'o}mez-Alvarez, Altieri, Marcum, Vavrek, Laureijs, Kohley, Buitrago, Cuillandre, Duc, {et~al.}}]{borlaff2022euclid}
Borlaff, A.~S., G{\'o}mez-Alvarez, P., Altieri, B., {et~al.} 2022, Astronomy \& Astrophysics, 657, A92

\bibitem[{Boselli {et~al.}(2016)Boselli, Cuillandre, Fossati, Boissier, Bomans, Consolandi, Anselmi, Cortese, C{\^o}t{\'e}, Durrell, {et~al.}}]{boselli2016spectacular}
Boselli, A., Cuillandre, J., Fossati, M., {et~al.} 2016, Astronomy \& Astrophysics, 587, A68

\bibitem[{Cao {et~al.}(2018)Cao, Gong, Meng, Xu, Chen, Guo, Li, Liu, Xue, Cao, {et~al.}}]{cao2018testing}
Cao, Y., Gong, Y., Meng, X.-M., {et~al.} 2018, Monthly Notices of the Royal Astronomical Society, 480, 2178

\bibitem[{Capaccioli \& Schipani(2011)}]{capaccioli2011vlt}
Capaccioli, M. \& Schipani, P. 2011, The Messenger, 146, 27

\bibitem[{{Chilingarian} {et~al.}(2019){Chilingarian}, {Afanasiev}, {Grishin}, {Fabricant}, \& {Moran}}]{Chilingarian2019_udg_spec}
{Chilingarian}, I.~V., {Afanasiev}, A.~V., {Grishin}, K.~A., {Fabricant}, D., \& {Moran}, S. 2019, \apj, 884, 79

\bibitem[{De~Blok \& McGaugh(1997)}]{de1997dark}
De~Blok, W. \& McGaugh, S.~S. 1997, Monthly Notices of the Royal Astronomical Society, 290, 533

\bibitem[{de~Jong {et~al.}(2015)de~Jong, Kleijn, Boxhoorn, Buddelmeijer, Capaccioli, Getman, Grado, Helmich, Huang, Irisarri, {et~al.}}]{de2015first}
de~Jong, J.~T., Kleijn, G. A.~V., Boxhoorn, D.~R., {et~al.} 2015, Astronomy \& Astrophysics, 582, A62

\bibitem[{De~Jong {et~al.}(2017)De~Jong, Kleijn, Erben, Hildebrandt, Kuijken, Sikkema, Brescia, Bilicki, Napolitano, Amaro, {et~al.}}]{de2017third}
De~Jong, J.~T., Kleijn, G. A.~V., Erben, T., {et~al.} 2017, Astronomy \& Astrophysics, 604, A134

\bibitem[{de~Jong {et~al.}(2013)de~Jong, Verdoes~Kleijn, Kuijken, Valentijn, KiDS, \& Consortiums}]{de2013kilo}
de~Jong, J.~T., Verdoes~Kleijn, G.~A., Kuijken, K.~H., {et~al.} 2013, Experimental Astronomy, 35, 25

\bibitem[{Di~Cintio {et~al.}(2024)Di~Cintio, Arjona-Galvez, \& Grand}]{di2024role}
Di~Cintio, A., Arjona-Galvez, E., \& Grand, R. 2024, EAS2024, 294

\bibitem[{Driver {et~al.}(2011)Driver, Hill, Kelvin, Robotham, Liske, Norberg, Baldry, Bamford, Hopkins, Loveday, {et~al.}}]{driver2011galaxy}
Driver, S.~P., Hill, D.~T., Kelvin, L.~S., {et~al.} 2011, Monthly Notices of the Royal Astronomical Society, 413, 971

\bibitem[{Du {et~al.}(2015)Du, Wu, Lam, Zhu, Lei, \& Zhou}]{du2015low}
Du, W., Wu, H., Lam, M.~I., {et~al.} 2015, The Astronomical Journal, 149, 199

\bibitem[{Edge {et~al.}(2013)Edge, Sutherland, Kuijken, Driver, McMahon, Eales, \& Emerson}]{edge2013vista}
Edge, A., Sutherland, W., Kuijken, K., {et~al.} 2013, The Messenger, 154, 32

\bibitem[{Erwin(2015)}]{erwin2015imfit}
Erwin, P. 2015, The Astrophysical Journal, 799, 226

\bibitem[{Forbes {et~al.}(2019)Forbes, Gannon, Couch, Iodice, Spavone, Cantiello, Napolitano, \& Schipani}]{forbes2019ultra}
Forbes, D.~A., Gannon, J., Couch, W.~J., {et~al.} 2019, Astronomy \& Astrophysics, 626, A66

\bibitem[{{Gannon} {et~al.}(2024){Gannon}, {Ferr{\'e}-Mateu}, {Forbes}, {Brodie}, {Buzzo}, \& {Romanowsky}}]{2024Gannon_spec_rev}
{Gannon}, J.~S., {Ferr{\'e}-Mateu}, A., {Forbes}, D.~A., {et~al.} 2024, \mnras, 531, 1856

\bibitem[{Gannon {et~al.}(2022)Gannon, Forbes, Romanowsky, Ferr{\'e}-Mateu, Couch, Brodie, Huang, Janssens, \& Okabe}]{gannon2022ultra}
Gannon, J.~S., Forbes, D.~A., Romanowsky, A.~J., {et~al.} 2022, Monthly Notices of the Royal Astronomical Society, 510, 946

\bibitem[{Gong {et~al.}(2019)Gong, Liu, Cao, Chen, Fan, Li, Li, Li, Zhang, \& Zhan}]{gong2019cosmology}
Gong, Y., Liu, X., Cao, Y., {et~al.} 2019, The Astrophysical Journal, 883, 203

\bibitem[{Gonz{\'a}lez {et~al.}(2018)Gonz{\'a}lez, Munoz, \& Hern{\'a}ndez}]{gonzalez2018galaxy}
Gonz{\'a}lez, R.~E., Munoz, R.~P., \& Hern{\'a}ndez, C.~A. 2018, Astronomy and computing, 25, 103

\bibitem[{Greco {et~al.}(2018)Greco, Greene, Strauss, Macarthur, Flowers, Goulding, Huang, Kim, Komiyama, Leauthaud, {et~al.}}]{greco2018illuminating}
Greco, J.~P., Greene, J.~E., Strauss, M.~A., {et~al.} 2018, The Astrophysical Journal, 857, 104

\bibitem[{Grishin {et~al.}(2023)Grishin, Mei, \& Ili{\'c}}]{grishin2023yolo}
Grishin, K., Mei, S., \& Ili{\'c}, S. 2023, Astronomy \& Astrophysics, 677, A101

\bibitem[{Hayward {et~al.}(2005)Hayward, Irwin, \& Bregman}]{hayward2005cosmological}
Hayward, C.~C., Irwin, J., \& Bregman, J. 2005, The Astrophysical Journal, 635, 827

\bibitem[{He {et~al.}(2020)He, Wu, Du, Liu, Lei, Zhao, \& Zhang}]{he2020sample}
He, M., Wu, H., Du, W., {et~al.} 2020, The Astrophysical Journal Supplement Series, 248, 33

\bibitem[{Hou {et~al.}(2021)Hou, Zhou, \& Feng}]{hou2021coordinate}
Hou, Q., Zhou, D., \& Feng, J. 2021, in Proceedings of the IEEE/CVF conference on computer vision and pattern recognition, 13713--13722

\bibitem[{Impey {et~al.}(1988)Impey, Bothun, \& Malin}]{impey1988virgo}
Impey, C., Bothun, G., \& Malin, D. 1988, Astrophysical Journal, Part 1 (ISSN 0004-637X), vol. 330, July 15, 1988, p. 634-660., 330, 634

\bibitem[{Iodice {et~al.}(2020)Iodice, Cantiello, Hilker, Rejkuba, Arnaboldi, Spavone, Greggio, Forbes, D’Ago, Mieske, {et~al.}}]{iodice2020first}
Iodice, E., Cantiello, M., Hilker, M., {et~al.} 2020, Astronomy \& Astrophysics, 642, A48

\bibitem[{Ivezi{\'c} {et~al.}(2019)Ivezi{\'c}, Kahn, Tyson, Abel, Acosta, Allsman, Alonso, AlSayyad, Anderson, Andrew, {et~al.}}]{ivezic2019lsst}
Ivezi{\'c}, {\v{Z}}., Kahn, S.~M., Tyson, J.~A., {et~al.} 2019, The Astrophysical Journal, 873, 111

\bibitem[{{Kelvin} {et~al.}(2018){Kelvin}, {Bremer}, {Phillipps}, {James}, {Davies}, {De Propris}, {Moffett}, {Percival}, {Baldry}, {Collins}, {Alpaslan}, {Bland-Hawthorn}, {Brough}, {Cluver}, {Driver}, {Hashemizadeh}, {Holwerda}, {Laine}, {Lara-Lopez}, {Liske}, {Maciejewski}, {Napolitano}, {Penny}, {Popescu}, {Sansom}, {Sutherland}, {Taylor}, {van Kampen}, \& {Wang}}]{2018kelvin_surface_bright_kids}
{Kelvin}, L.~S., {Bremer}, M.~N., {Phillipps}, S., {et~al.} 2018, \mnras, 477, 4116

\bibitem[{Koda {et~al.}(2015)Koda, Yagi, Yamanoi, \& Komiyama}]{koda2015approximately}
Koda, J., Yagi, M., Yamanoi, H., \& Komiyama, Y. 2015, The Astrophysical Journal Letters, 807, L2

\bibitem[{Kuijken {et~al.}(2019)Kuijken, Heymans, Dvornik, Hildebrandt, de~Jong, Wright, Erben, Bilicki, Giblin, Shan, {et~al.}}]{kuijken2019fourth}
Kuijken, K., Heymans, C., Dvornik, A., {et~al.} 2019, Astronomy \& Astrophysics, 625, A2

\bibitem[{Kuijken {et~al.}(2011)}]{kuijken2011omegacam}
Kuijken, K. {et~al.} 2011, The Messenger, 146

\bibitem[{La~Marca {et~al.}(2022)La~Marca, Iodice, Cantiello, Forbes, Rejkuba, Hilker, Arnaboldi, Greggio, Spiniello, Mieske, {et~al.}}]{la2022galaxy}
La~Marca, A., Iodice, E., Cantiello, M., {et~al.} 2022, Astronomy \& Astrophysics, 665, A105

\bibitem[{Laureijs {et~al.}(2011)Laureijs, Amiaux, Arduini, Augueres, Brinchmann, Cole, Cropper, Dabin, Duvet, Ealet, {et~al.}}]{laureijs2011euclid}
Laureijs, R., Amiaux, J., Arduini, S., {et~al.} 2011, arXiv preprint arXiv:1110.3193

\bibitem[{Lee {et~al.}(2017)Lee, Kang, Lee, \& Jang}]{lee2017detection}
Lee, M.~G., Kang, J., Lee, J.~H., \& Jang, I.~S. 2017, The Astrophysical Journal, 844, 157

\bibitem[{Leisman {et~al.}(2017)Leisman, Haynes, Janowiecki, Hallenbeck, J{\'o}zsa, Giovanelli, Adams, Neira, Cannon, Janesh, {et~al.}}]{leisman2017almost}
Leisman, L., Haynes, M.~P., Janowiecki, S., {et~al.} 2017, The Astrophysical Journal, 842, 133

\bibitem[{Lim {et~al.}(2020)Lim, C{\^o}t{\'e}, Peng, Ferrarese, Roediger, Durrell, Mihos, Wang, Gwyn, Cuillandre, {et~al.}}]{lim2020next}
Lim, S., C{\^o}t{\'e}, P., Peng, E.~W., {et~al.} 2020, The Astrophysical Journal, 899, 69

\bibitem[{Liu {et~al.}(2017)Liu, Jiang, Faber, Koo, Yesuf, Tacchella, Mao, Wang, Guo, Fang, {et~al.}}]{liu2017origins}
Liu, F., Jiang, D., Faber, S.~M., {et~al.} 2017, The Astrophysical Journal Letters, 844, L2

\bibitem[{Marleau {et~al.}(2025)Marleau, Habas, Carollo, Tortora, Duc, Sola, Saifollahi, F{\"u}genschuh, Walmsley, Z{\"o}ller, {et~al.}}]{marleau2025euclid}
Marleau, F., Habas, R., Carollo, D., {et~al.} 2025, arXiv preprint arXiv:2503.15335

\bibitem[{{Marleau} {et~al.}(2021){Marleau}, {Habas}, {Poulain}, {Duc}, {M{\"u}ller}, {Lim}, {Durrell}, {S{\'a}nchez-Janssen}, {Paudel}, {Ahad}, {Chougule}, {B{\'\i}lek}, \& {Fensch}}]{2021MATLAS_UDG}
{Marleau}, F.~R., {Habas}, R., {Poulain}, M., {et~al.} 2021, \aap, 654, A105

\bibitem[{Marleau {et~al.}(2021)Marleau, Habas, Poulain, Duc, M{\"u}ller, Lim, Durrell, Sanchez-Janssen, Paudel, Ahad, {et~al.}}]{marleau2021ultra}
Marleau, F.~R., Habas, R., Poulain, M., {et~al.} 2021, Astronomy \& Astrophysics, 654, A105

\bibitem[{Merritt {et~al.}(2016)Merritt, van Dokkum, Danieli, Abraham, Zhang, Karachentsev, \& Makarova}]{merritt2016dragonfly}
Merritt, A., van Dokkum, P., Danieli, S., {et~al.} 2016, The Astrophysical Journal, 833, 168

\bibitem[{Mihos {et~al.}(2016)Mihos, Harding, Feldmeier, Rudick, Janowiecki, Morrison, Slater, \& Watkins}]{mihos2016burrell}
Mihos, J.~C., Harding, P., Feldmeier, J.~J., {et~al.} 2016, The Astrophysical Journal, 834, 16

\bibitem[{Peng {et~al.}(2002)Peng, Ho, Impey, \& Rix}]{peng2002detailed}
Peng, C.~Y., Ho, L.~C., Impey, C.~D., \& Rix, H.-W. 2002, The Astronomical Journal, 124, 266

\bibitem[{Pina {et~al.}(2019)Pina, Fraternali, Adams, Marasco, Oosterloo, Oman, Leisman, Di~Teodoro, Posti, Battipaglia, {et~al.}}]{pina2019off}
Pina, P. E.~M., Fraternali, F., Adams, E.~A., {et~al.} 2019, The Astrophysical Journal Letters, 883, L33

\bibitem[{{Planck Collaboration} {et~al.}(2020){Planck Collaboration}, {Aghanim}, {Akrami}, {Ashdown}, {Aumont}, {Baccigalupi}, {Ballardini}, {Banday}, {Barreiro}, {Bartolo}, {Basak}, {Battye}, {Benabed}, {Bernard}, {Bersanelli}, {Bielewicz}, {Bock}, {Bond}, {Borrill}, {Bouchet}, {Boulanger}, {Bucher}, {Burigana}, {Butler}, {Calabrese}, {Cardoso}, {Carron}, {Challinor}, {Chiang}, {Chluba}, {Colombo}, {Combet}, {Contreras}, {Crill}, {Cuttaia}, {de Bernardis}, {de Zotti}, {Delabrouille}, {Delouis}, {Di Valentino}, {Diego}, {Dor{\'e}}, {Douspis}, {Ducout}, {Dupac}, {Dusini}, {Efstathiou}, {Elsner}, {En{\ss}lin}, {Eriksen}, {Fantaye}, {Farhang}, {Fergusson}, {Fernandez-Cobos}, {Finelli}, {Forastieri}, {Frailis}, {Fraisse}, {Franceschi}, {Frolov}, {Galeotta}, {Galli}, {Ganga}, {G{\'e}nova-Santos}, {Gerbino}, {Ghosh}, {Gonz{\'a}lez-Nuevo}, {G{\'o}rski}, {Gratton}, {Gruppuso}, {Gudmundsson}, {Hamann}, {Handley}, {Hansen}, {Herranz}, {Hildebrandt}, {Hivon}, {Huang}, {Jaffe}, {Jones}, {Karakci}, {Keih{\"a}nen},
  {Keskitalo}, {Kiiveri}, {Kim}, {Kisner}, {Knox}, {Krachmalnicoff}, {Kunz}, {Kurki-Suonio}, {Lagache}, {Lamarre}, {Lasenby}, {Lattanzi}, {Lawrence}, {Le Jeune}, {Lemos}, {Lesgourgues}, {Levrier}, {Lewis}, {Liguori}, {Lilje}, {Lilley}, {Lindholm}, {L{\'o}pez-Caniego}, {Lubin}, {Ma}, {Mac{\'\i}as-P{\'e}rez}, {Maggio}, {Maino}, {Mandolesi}, {Mangilli}, {Marcos-Caballero}, {Maris}, {Martin}, {Martinelli}, {Mart{\'\i}nez-Gonz{\'a}lez}, {Matarrese}, {Mauri}, {McEwen}, {Meinhold}, {Melchiorri}, {Mennella}, {Migliaccio}, {Millea}, {Mitra}, {Miville-Desch{\^e}nes}, {Molinari}, {Montier}, {Morgante}, {Moss}, {Natoli}, {N{\o}rgaard-Nielsen}, {Pagano}, {Paoletti}, {Partridge}, {Patanchon}, {Peiris}, {Perrotta}, {Pettorino}, {Piacentini}, {Polastri}, {Polenta}, {Puget}, {Rachen}, {Reinecke}, {Remazeilles}, {Renzi}, {Rocha}, {Rosset}, {Roudier}, {Rubi{\~n}o-Mart{\'\i}n}, {Ruiz-Granados}, {Salvati}, {Sandri}, {Savelainen}, {Scott}, {Shellard}, {Sirignano}, {Sirri}, {Spencer}, {Sunyaev}, {Suur-Uski}, {Tauber}, {Tavagnacco},
  {Tenti}, {Toffolatti}, {Tomasi}, {Trombetti}, {Valenziano}, {Valiviita}, {Van Tent}, {Vibert}, {Vielva}, {Villa}, {Vittorio}, {Wandelt}, {Wehus}, {White}, {White}, {Zacchei}, \& {Zonca}}]{Planck2020}
{Planck Collaboration}, {Aghanim}, N., {Akrami}, Y., {et~al.} 2020, \aap, 641, A6

\bibitem[{Prole {et~al.}(2019{\natexlab{a}})Prole, van~der Burg, Hilker, \& Davies}]{prole2019observational}
Prole, D., van~der Burg, R., Hilker, M., \& Davies, J. 2019{\natexlab{a}}, Monthly Notices of the Royal Astronomical Society, 488, 2143

\bibitem[{Prole {et~al.}(2019{\natexlab{b}})Prole, Hilker, van~der Burg, Cantiello, Venhola, Iodice, van~de Ven, Wittmann, Peletier, Mieske, {et~al.}}]{prole2019halo}
Prole, D.~J., Hilker, M., van~der Burg, R.~F., {et~al.} 2019{\natexlab{b}}, Monthly Notices of the Royal Astronomical Society, 484, 4865

\bibitem[{Redmon(2016)}]{redmon2016you}
Redmon, J. 2016, in Proceedings of the IEEE conference on computer vision and pattern recognition

\bibitem[{Rom{\'a}n \& Trujillo(2017{\natexlab{a}})}]{roman2017spatial}
Rom{\'a}n, J. \& Trujillo, I. 2017{\natexlab{a}}, Monthly Notices of the Royal Astronomical Society, 468, 703

\bibitem[{Rom{\'a}n \& Trujillo(2017{\natexlab{b}})}]{roman2017ultra}
Rom{\'a}n, J. \& Trujillo, I. 2017{\natexlab{b}}, Monthly Notices of the Royal Astronomical Society, 468, 4039

\bibitem[{{Roy} {et~al.}(2018){Roy}, {Napolitano}, {La Barbera}, {Tortora}, {Getman}, {Radovich}, {Capaccioli}, {Brescia}, {Cavuoti}, {Longo}, {Raj}, {Puddu}, {Covone}, {Amaro}, {Vellucci}, {Grado}, {Kuijken}, {Verdoes Kleijn}, \& {Valentijn}}]{2018roy_kids_sersic}
{Roy}, N., {Napolitano}, N.~R., {La Barbera}, F., {et~al.} 2018, \mnras, 480, 1057

\bibitem[{Saifollahi {et~al.}(2022)Saifollahi, Zaritsky, Trujillo, Peletier, Knapen, Amorisco, Beasley, \& Donnerstein}]{saifollahi2022implications}
Saifollahi, T., Zaritsky, D., Trujillo, I., {et~al.} 2022, Monthly Notices of the Royal Astronomical Society, 511, 4633

\bibitem[{Sevilla-Noarbe {et~al.}(2021)Sevilla-Noarbe, Bechtol, Kind, Rosell, Becker, Drlica-Wagner, Gruendl, Rykoff, Sheldon, Yanny, {et~al.}}]{sevilla2021dark}
Sevilla-Noarbe, I., Bechtol, K., Kind, M.~C., {et~al.} 2021, The Astrophysical Journal Supplement Series, 254, 24

\bibitem[{Su {et~al.}(2024)Su, Yi, Liang, Du, Liu, Kong, Bu, \& Wu}]{su2024lsbgnet}
Su, H., Yi, Z., Liang, Z., {et~al.} 2024, Monthly Notices of the Royal Astronomical Society, 528, 873

\bibitem[{Tanoglidis {et~al.}(2021)Tanoglidis, Drlica-Wagner, Wei, Li, S{\'a}nchez, Zhang, Peter, Feldmeier-Krause, Prat, Casey, {et~al.}}]{tanoglidis2021shadows}
Tanoglidis, D., Drlica-Wagner, A., Wei, K., {et~al.} 2021, The Astrophysical Journal Supplement Series, 252, 18

\bibitem[{Teeninga {et~al.}(2016)Teeninga, Moschini, Trager, \& Wilkinson}]{teeninga2016statistical}
Teeninga, P., Moschini, U., Trager, S.~C., \& Wilkinson, M.~H. 2016, Mathematical Morphology-Theory and Applications, 1

\bibitem[{{Thuruthipilly} {et~al.}(2025){Thuruthipilly}, {Junais}, {Koda}, {Pollo}, {Yagi}, {Yamanoi}, {Komiyama}, {Romano}, {Ma{\l}ek}, \& {Donevski}}]{2025_Thuruthipilly_fluxrad}
{Thuruthipilly}, H., {Junais}, {Koda}, J., {et~al.} 2025, \aap, 695, A106

\bibitem[{Toloba {et~al.}(2015)Toloba, Sand, Spekkens, Crnojevi{\'c}, Simon, Guhathakurta, Strader, Caldwell, McLeod, \& Seth}]{toloba2015tidally}
Toloba, E., Sand, D.~J., Spekkens, K., {et~al.} 2015, The Astrophysical Journal Letters, 816, L5

\bibitem[{Tremmel {et~al.}(2020)Tremmel, Wright, Brooks, Munshi, Nagai, \& Quinn}]{tremmel2020formation}
Tremmel, M., Wright, A.~C., Brooks, A.~M., {et~al.} 2020, Monthly Notices of the Royal Astronomical Society, 497, 2786

\bibitem[{Tzutalin(2015)}]{tzutalin2015labelimg}
Tzutalin, D. 2015, GitHub repository, 6, 4

\bibitem[{van Der~Burg {et~al.}(2017)van Der~Burg, Hoekstra, Muzzin, Sif{\'o}n, Viola, Bremer, Brough, Driver, Erben, Heymans, {et~al.}}]{van2017abundance}
van Der~Burg, R.~F., Hoekstra, H., Muzzin, A., {et~al.} 2017, Astronomy \& Astrophysics, 607, A79

\bibitem[{Van~Dokkum {et~al.}(2015)Van~Dokkum, Abraham, Merritt, Zhang, Geha, \& Conroy}]{van2015forty}
Van~Dokkum, P.~G., Abraham, R., Merritt, A., {et~al.} 2015, The Astrophysical Journal Letters, 798, L45

\bibitem[{Venemans {et~al.}(2015)Venemans, Verdoes~Kleijn, Mwebaze, Valentijn, Banados, Decarli, de~Jong, Findlay, Kuijken, Barbera, {et~al.}}]{venemans2015first}
Venemans, B., Verdoes~Kleijn, G., Mwebaze, J., {et~al.} 2015, Monthly Notices of the Royal Astronomical Society, 453, 2259

\bibitem[{Venhola {et~al.}(2018)Venhola, Peletier, Laurikainen, Salo, Iodice, Mieske, Hilker, Wittmann, Lisker, Paolillo, {et~al.}}]{venhola2018fornax}
Venhola, A., Peletier, R., Laurikainen, E., {et~al.} 2018, Astronomy \& Astrophysics, 620, A165

\bibitem[{Venhola {et~al.}(2017)Venhola, Peletier, Laurikainen, Salo, Lisker, Iodice, Capaccioli, Kleijn, Valentijn, Mieske, {et~al.}}]{venhola2017fornax}
Venhola, A., Peletier, R., Laurikainen, E., {et~al.} 2017, Astronomy \& Astrophysics, 608, A142

\bibitem[{Wittmann {et~al.}(2017)Wittmann, Lisker, Ambachew~Tilahun, Grebel, Conselice, Penny, Janz, Gallagher~III, Kotulla, \& McCormac}]{wittmann2017population}
Wittmann, C., Lisker, T., Ambachew~Tilahun, L., {et~al.} 2017, Monthly Notices of the Royal Astronomical Society, 470, 1512

\bibitem[{Wright {et~al.}(2024)Wright, Kuijken, Hildebrandt, Radovich, Bilicki, Dvornik, Getman, Heymans, Hoekstra, Li, {et~al.}}]{wright2024fifth}
Wright, A.~H., Kuijken, K., Hildebrandt, H., {et~al.} 2024, Astronomy \& Astrophysics, 686, A170

\bibitem[{Yagi {et~al.}(2016)Yagi, Koda, Komiyama, \& Yamanoi}]{yagi2016catalog}
Yagi, M., Koda, J., Komiyama, Y., \& Yamanoi, H. 2016, The Astrophysical Journal Supplement Series, 225, 11

\bibitem[{Yi {et~al.}(2022)Yi, Li, Du, Liu, Liang, Xing, Pan, Bu, Kong, \& Wu}]{yi2022automatic}
Yi, Z., Li, J., Du, W., {et~al.} 2022, Monthly Notices of the Royal Astronomical Society, 513, 3972

\bibitem[{Yozin \& Bekki(2015)}]{yozin2015quenching}
Yozin, C. \& Bekki, K. 2015, Monthly Notices of the Royal Astronomical Society, 452, 937

\bibitem[{Zaritsky {et~al.}(2018)Zaritsky, Donnerstein, Dey, Kadowaki, Zhang, Karunakaran, Mart{\'\i}nez-Delgado, Rahman, \& Spekkens}]{zaritsky2018systematically}
Zaritsky, D., Donnerstein, R., Dey, A., {et~al.} 2018, The Astrophysical Journal Supplement Series, 240, 1

\bibitem[{Zaritsky {et~al.}(2023)Zaritsky, Donnerstein, Dey, Karunakaran, Kadowaki, Khim, Spekkens, \& Zhang}]{zaritsky2023systematically}
Zaritsky, D., Donnerstein, R., Dey, A., {et~al.} 2023, The Astrophysical Journal Supplement Series, 267, 27

\bibitem[{Zaritsky {et~al.}(2021)Zaritsky, Donnerstein, Karunakaran, Barbosa, Dey, Kadowaki, Spekkens, \& Zhang}]{zaritsky2021systematically}
Zaritsky, D., Donnerstein, R., Karunakaran, A., {et~al.} 2021, The Astrophysical Journal Supplement Series, 257, 60

\bibitem[{Zhan(2011)}]{zhan2011consideration}
Zhan, H. 2011, Scientia Sinica Physica, Mechanica \& Astronomica, 41, 1441

\end{thebibliography}
\bibliographystyle{aa}
\begin{appendix}
\begin{sidewaystable*}
\section{List of UDGs}
\centering
\caption{Catalog of High-Quality UDG Candidates in matched clusters/groups}
\label{apppendix}
\begin{tabular}{lccccccccc}
\hline\hline
ID & Cluster & RA (deg) & Dec (deg) & $\langle \mu_e \rangle_r$ (mag/arcsec$^2$) & Distance (Mpc) & $R_e$ (kpc) & mean\_score & $ \sigma $ & $\langle {\rm Grade} \rangle$ \\
\hline
KiDS\_UDG\_Fornax\_51.72\_35.51            & Fornax           & 51.7249   & -35.5178 & 24.79 & 26.8     & 2.029  &10.0 &0.0 &A\\
KiDS\_UDG\_Fornax\_52.11\_33.73            & Fornax           & 52.1181   & -33.7348 & 24.35  & 26.8     & 1.614 &10.0 &0.0&A\\
KiDS\_UDG\_Fornax\_53.58\_35.36            & Fornax           & 53.5841   & -35.3624 & 24.44 & 26.8     & 1.456  &10.0 &0.0&A\\
KiDS\_UDG\_Fornax\_53.92\_35.51            & Fornax           & 53.9281   & -35.5142 & 25.35  & 26.8    & 1.743  &9.14 &1.46&A\\
KiDS\_UDG\_Fornax\_53.98\_34.82             & Fornax           & 53.982    & -34.8283 & 25.21 & 26.8     & 1.863  &10.0 &0.0&A\\
KiDS\_UDG\_IC1860\_41.93\_31.05           & IC 1860          & 41.9304   & -31.0538 & 25.26 & 96.5     & 3.212  &9.57 &1.13&A\\
KiDS\_UDG\_IC1860\_42.19\_31.52            & IC 1860          & 42.191    & -31.5262 & 24.90 & 96.5     & 2.011  &9.14 &1.46&A\\
KiDS\_UDG\_IC5157\_330.40\_35.22          & IC 5157          & 330.4094  & -35.2232 & 24.16 & 64.1     & 1.743  &8.8 &1.64&A\\
KiDS\_UDG\_IC5157\_330.70\_34.75          & IC 5157          & 330.7037  & -34.7567 & 24.99 & 64.1     & 1.462  &8.71 &1.6&A\\
KiDS\_UDG\_IC5157\_330.72\_34.90             & IC 5157          & 330.7258  & -34.9    & 24.14 & 64.1     & 1.879  &9.4 &1.34&A\\
KiDS\_UDG\_IC5157\_330.96\_35.20          & IC 5157          & 330.9632  & -35.2004 & 24.48 & 64.1     & 1.325  &9 &1.55&A\\
KiDS\_UDG\_IC5358\_356.70\_28.12            & IC 5358          & 356.706   & -28.124  & 24.87 & 123.0     & 2.582  &9.25 &1.39&A\\
KiDS\_UDG\_IC5358\_356.78\_28.24          & IC 5358          & 356.7852  & -28.2484 & 24.18 & 123.0    & 3.567  &8.8 &1.64&A\\
KiDS\_UDG\_NGC0439\_18.56\_32.48          & NGC 0439         & 18.5622   & -32.4841 & 24.63 & 83.1     & 1.839  &9.4 &1.34&A\\
KiDS\_UDG\_NGC0439\_18.60\-32.46           & NGC 0439         & 18.6007   & -32.462  & 25.77 & 83.1     & 1.74   &9.57 &1.13&A\\
KiDS\_UDG\_NGC0439\_18.63\_32.19          & NGC 0439         & 18.6397   & -32.1925 & 24.99 & 83.1    & 1.702  &7.14 &3.13&A\\
KiDS\_UDG\_NGC0439\_18.67\_32.20          & NGC 0439         & 18.6717   & -32.2038 & 24.26 & 83.1     & 3.515  &10.0 &0.0&A\\
KiDS\_UDG\_NGC0439\_19.01\_31.76          & NGC 0439         & 19.0112   & -31.7601 & 24.57 & 83.1     & 1.733  &8.12 &1.55&A\\
KiDS\_UDG\_NGC4073\_180.58\_2.26           & NGC 4073         & 180.5874  & 2.2665   & 23.86 & 87.5     & 2.152  &9.5 &1.22&A\\
KiDS\_UDG\_NGC4073\_180.66\_1.83           & NGC 4073         & 180.6606  & 1.8314   & 24.41 & 87.5     & 1.888  &9.0 &1.55&A\\
KiDS\_UDG\_NGC4073\_180.84\_1.79             & NGC 4073         & 180.8461  & 1.79     & 25.31 & 87.5     & 1.916  &8.57 &2.7&A\\
KiDS\_UDG\_NGC4073\_180.86\_1.40           & NGC 4073         & 180.8698  & 1.4058   & 24.05  & 87.5     & 2.216  &10.0 &0.0&A\\
KiDS\_UDG\_NGC4073\_180.88\_1.80            & NGC 4073         & 180.8865  & 1.803    & 23.99 & 87.5     & 2.64   &10.0 &0.0&A\\
KiDS\_UDG\_NGC4073\_180.92\_1.93           & NGC 4073         & 180.9242  & 1.9327   & 25.62 & 87.5     & 2.296  &9.57 &1.13&A\\
KiDS\_UDG\_NGC4073\_181.03\_1.58           & NGC 4073         & 181.0309  & 1.5832   & 24.55  & 87.5     & 2.784  &10.0 &0.0&A\\
KiDS\_UDG\_NGC4073\_181.03\_1.32           & NGC 4073         & 181.0329  & 1.3216   & 25.40 & 87.5     & 1.556  &8.71 &1.6&A\\
KiDS\_UDG\_NGC4073\_181.04\_1.87           & NGC 4073         & 181.0409  & 1.8736   & 24.41 & 87.5     & 1.888  &10.0 &0.0&A\\
KiDS\_UDG\_NGC4073\_181.13\_1.41            & NGC 4073         & 181.1397  & 1.416    & 25.36 & 87.5     & 1.804  &9.14 &1.46&A\\
KiDS\_UDG\_NGC4073\_181.17\_1.45             & NGC 4073         & 181.17    & 1.4507   & 25.14 & 87.5     & 1.72   &8.71 &1.6&A\\
KiDS\_UDG\_NGC4073\_181.24\_1.98            & NGC 4073         & 181.247   & 1.9873   & 24.18 & 87.5    & 1.888  &8.8 &1.64&A\\
KiDS\_UDG\_NGC4073\_181.24\_1.95           & NGC 4073         & 181.2471  & 1.9554   & 24.94 & 87.5    & 1.86   &7.29 &3.55&A\\
KiDS\_UDG\_NGC4073\_181.35\_1.92            & NGC 4073         & 181.352   & 1.9237   & 25.19 & 87.5     & 2.176  &9.14 &1.46&A\\
KiDS\_UDG\_NGC4073\_181.78\_1.90           & NGC 4073         & 181.7882  & 1.9005   & 25.05  & 87.5     & 1.824  &9.14 &1.46&A\\
KiDS\_UDG\_NGC4636\_188.23\_3.36           & NGC 4636         & 188.2329  & 3.3605   & 23.92 & 17.5     & 1.682  &8.8 &1.64&A\\
KiDS\_UDG\_NGC4636\_190.46\_2.74            & NGC 4636         & 190.466   & 2.7446   & 24.06 & 17.5     & 1.516  &10.0 &0.0&A\\
KiDS\_UDG\_NGC4636\_191.23\_2.29           & NGC 4636         & 191.2318  & 2.2959   & 24.67 & 17.5     & 1.414  &10.0 &0.0 &A\\
KiDS\_UDG\_NGC4636\_192.50\_3.21           & NGC 4636         & 192.5062  & 3.2154   & 25.40 & 17.5     & 1.678  &9.14 &1.46&A\\
KiDS\_UDG\_NGC4636\_191.82\_1.72           & NGC 4636         & 191.82201 & -1.7203  & 26.10 & 17.5     & 1.412  &9.14 &1.46&A\\
KiDS\_UDG\_NGC5576\_191.84\_1.40          & NGC 5576         & 191.8478  & -1.4049  & 25.90 & 22.5     & 1.896  &9.57 &1.13&A\\
KiDS\_UDG\_NGC5576\_191.98\_2.54          & NGC 5576         & 191.9899  & -2.5431  & 25.12 & 22.5    & 1.649  &8.38 &2.56&A\\
KiDS\_UDG\_NGC5576\_192.05\_0.83          & NGC 5576         & 192.0504  & -0.8391  & 24.12 & 22.5     & 1.412  &8.8 &1.64&A\\
\hline
\end{tabular}
\end{sidewaystable*}

\begin{sidewaystable*}
\centering
\caption{continued}
\begin{tabular}{lccccccccc}
\hline\hline
ID & Cluster & RA (deg) & Dec (deg) & $\langle \mu_e \rangle_r$ (mag/arcsec$^2$) & Distance (Mpc) & $R_e$ (kpc) & mean\_score &$ \sigma $& $\langle {\rm Grade} \rangle$ \\
\hline
KiDS\_UDG\_NGC5576\_192.23\_0.78            & NGC 5576         & 192.2392  & -0.78    & 25.65  & 22.5     & 1.4    &9.14 &1.46&A\\
KiDS\_UDG\_NGC4690\_214.81\_3.12           & NGC 4690         & 214.8123  & 3.1243   & 25.13 & 40.1     & 1.568  &10.0 &0.0&A\\
KiDS\_UDG\_NGC4690\_217.02\_3.18            & NGC 4690         & 217.029   & 3.1863   & 26.13 & 40.1     & 1.447  &9.57 &1.13&A\\
KiDS\_UDG\_NGC4690\_215.51\_3.58           & NGC 4690         & 215.5182  & 3.5866   & 25.62 & 40.1     & 1.642  &9.57 &1.13&A\\
KiDS\_UDG\_NGC4690\_217.71\_3.20          & NGC 4690         & 217.7127  & 3.2068   & 24.80 & 40.1     & 1.568  &10.0 &0.0&A\\
KiDS\_UDG\_NGC5846\_223.85\_1.03             & NGC 5846         & 223.8568  & 1.03     & 26.12 & 26.8     & 1.739 &9.57 &1.13&A \\
KiDS\_UDG\_NGC5846\_224.61\_1.54           & NGC 5846         & 224.6193  & 1.5422   & 23.96 & 26.8     & 1.434  &10.0 &0.0&A\\
KiDS\_UDG\_NGC5846\_225.83\_1.43           & NGC 5846         & 225.8391  & 1.4308   & 25.92 & 26.8     & 1.576  &8.71 &1.6&A\\
KiDS\_UDG\_NGC5846\_226.07\_1.75            & NGC 5846         & 226.0799  & 1.755    & 24.72 & 26.8     & 1.593  &8.5 &1.73&A\\
KiDS\_UDG\_NGC5846\_226.33\_1.81           & NGC 5846         & 226.3342  & 1.8115   & 25.40 & 26.8     & 2.267 &8.71 &1.6&A \\
KiDS\_UDG\_NGC7176\_330.10\_31.78         & NGC 7176         & 330.1037  & -31.786  & 23.90 & 36.4     & 1.577 &8.5 &1.73&A \\
KiDS\_UDG\_NGC7176\_330.56\_32.84         & NGC 7176         & 330.5658  & -32.8489 & 25.43 & 36.4     & 1.408  &8.71 &1.6&A\\
KiDS\_UDG\_NGC7176\_330.58\_31.57         & NGC 7176         & 330.5837  & -31.5786 & 24.40 & 36.4     & 1.415  &9.5 &1.22&A\\
KiDS\_UDG\_NGC7176\_331.36\_31.55         & NGC 7176         & 331.3604  & -31.5593 & 24.03 & 36.4     & 1.404  &10.0 &0.0&A\\
KiDS\_UDG\_NGC7176\_331.79\_31.07         & NGC 7176         & 331.7972  & -31.0796 & 24.20 & 36.4     & 1.818  &7.4 &2.88&A\\
KiDS\_UDG\_NGC7176\_331.81\_32.20          & NGC 7176         & 331.8153  & -32.207  & 24.08 & 36.4     & 1.528 &7.4 &2.88&A \\
KiDS\_UDG\_NGC7507\_347.06\_28.18          & NGC 7507         & 347.0601  & -28.183  & 24.97 & 22.9     & 1.457  &9.5 &1.22&A\\
KiDS\_UDG\_NGC7507\_347.969\_28.92          & NGC 7507         & 347.9699  & -28.926  & 25.23 & 22.9     & 1.513  &10.0 &0.0&A\\
KiDS\_UDG\_UGC07813\_189.78\_0.14          & UGC 07813        & 189.7877  & 0.1446   & 25.97 & 100.8     & 3.919  &9.57 &1.13&A\\
KiDS\_UDG\_UGC07813\_189.91\_0.82          & UGC 07813        & 189.9156  & 0.8249   & 25.09 & 100.8     & 2.778  &8.29 &1.6&A\\
KiDS\_UDG\_UGC07813\_189.94\_0.04          & UGC 07813        & 189.9486  & 0.0408   & 25.25  & 100.8     & 4.766  &10.0 &0.0&A\\
KiDS\_UDG\_UGC07813\_190.20\_0.44          & UGC 07813        & 190.2022  & 0.4496   & 23.86 & 100.8     & 3.427  &8.8 &1.64&A\\
KiDS\_UDG\_ESO466-021\_329.75\_28.44       & ESO466-021       & 329.7547  & -28.4454 & 24.44 & 100.8     & 3.096  &8.8 &1.64&A\\
KiDS\_UDG\_2MASSJ2259\_344.33\_34.09 & 2MASSJ2259-3334 & 344.3342  & -34.0913 & 25.78 & 123.0     & 4.334  &9.57 &1.13&A\\
KiDS\_UDG\_2MASSJ2259\_344.56\_33.68  & 2MASSJ2259-3334 & 344.5647  & -33.689  & 24.01 & 123.0     & 5.466  &10.0 &0.0&A\\
KiDS\_UDG\_2MASSJ2259\_345.68\_33.61 & 2MASSJ2259-3334 & 345.6859  & -33.6119 & 25.26 & 123.0     & 3.002 &8.8 &1.64&A\\
\hline
\end{tabular}
\end{sidewaystable*}

\end{appendix}
\end{document}